\documentclass[times,sort&compress,3p]{elsarticle}
\usepackage{graphicx,bm}
\usepackage{mathrsfs,amsmath,amsfonts}
\usepackage{multirow}
\usepackage{natbib}
\usepackage{color}
\usepackage{tcolorbox}
\usepackage[utf8x]{inputenc}
\usepackage{graphicx,subfigure}
\newtheorem{theorem}{Theorem}
\newtheorem{lemma}[theorem]{Lemma}

\numberwithin{equation}{section}
\begin{document}

\begin{frontmatter}

\title{An RKHS-Based Semiparametric Approach to Nonlinear Sufficient Dimension Reduction}
\author{Wenquan Cui and Haoyang Cheng\\\bigskip Department of Statistics and Finance\\The School of Management \\University of Science and Technology of China}
\date{June 2018}

\begin{abstract}
  Based on the theory of reproducing kernel Hilbert space (RKHS) and semiparametric method, we propose a new approach to nonlinear dimension reduction. The method extends the semiparametric method into a more generalized domain where both the interested parameters and nuisance parameters to be infinite dimensional. By casting the nonlinear dimensional reduction problem in a generalized semiparametric framework, we calculate the orthogonal complement space of generalized nuisance tangent space to derive the estimating equation. Solving the estimating equation by the theory of RKHS and regularization, we obtain the estimation of dimension reduction directions of the sufficient dimension reduction (SDR) subspace and also show the asymptotic property of estimator. Furthermore, the proposed method does not rely on the linearity condition and constant variance condition. Simulation and real data studies are conducted to demonstrate the finite sample performance of our method in comparison with several existing methods.
\end{abstract}

\begin{keyword}
  reproducing kernel, semiparametric methods, nuisance tangent space, sliced inverse regression, nonlinear dimension reduction
\end{keyword}

\end{frontmatter}

\section{Introduction}
Dimension reduction has been one of the hot research topics in the field of statistics and machine learning recently. With the data of many areas becoming more and more complex, statistical methods have to face the challenges of high dimension. Although high dimensional data contains a large amount of information in which people are interested, direct processing of high dimensional data is very difficult and tricky. Therefore, how to transform high dimensional data into a low dimensional model that can be processed with as little information loss as possible is significant. Currently, under the framework of supervised learning, the dimensional reduction methods for processing high dimensional data with sparse structures can be divided into two major categories - variable selection and sufficient dimensionality reduction (SDR). Assuming that the response variable Y is only related to few covariates $X_0$ and the remaining covariates $X_1$ have no effect on the response variables, the purpose of variable selection is to select these few important variables to reduce the data dimension. On the other hand, the goal of sufficient dimension reduction is to find the directions $\bm{\beta}$, so that Y is only related to the linear combination $\bm{\beta}^{T}X$. Assuming that the response variable Y depends on the $q~\left(q < p\right)$ linear combinations of the p-dimensional covariate X, there exist q numbers of p-dimensional vectors $\bm{\beta}_1, \bm{\beta}_2, \ldots, \bm{\beta}_q$. When given $X^T(\bm{\beta}_1, \bm{\beta}_2, \ldots, \bm{\beta}_q) = X^T\bm{\beta}$, Y and X are mutually independent, i.e.
\begin{equation}\label{1.1}
  F(y|X) = F(y|X^{T}\bm{\beta}), \forall y \in \mathbb{R}
\end{equation}
where $F(y|\cdot) = Pr(Y \le y | \cdot)$ is the conditional distribution of Y. The space that composed by $\bm{\beta}$ is called dimension reduction subspace. As dimension reduction subspace is not unique, we focus on the intersection of all dimension reduction subspaces, which is called the central subspace. We denote by $S_{Y|X} = span\{\bm{\beta}_1, \bm{\beta}_2, \ldots, \bm{\beta}_q\}$ the SDR central subspace.
our goal is to find $S_{Y|X}$ throughout finding $\bm{\beta}$ which satisfied (\ref{1.1}), which has been discussed in \cite{Cook1998, CookandLi2002}.

The sufficient dimension reduction based on the model (\ref{1.1}) obtains the dimensionality reduction features with a linear structure. To identify the SDR central subspace, there have been many studies on model (\ref{1.1}). Li \cite{Li1991} proposed the sliced inverse regression (SIR), which pioneered a number of "inverse regression" approaches, such as sliced average variance estimation (SAVE)\cite{Cook1991}, kernel inverse regression\cite{zhu1996asymptotics}, direction regression (DR)\cite{LiandWang2007}, etc. Besides the "inverse regression" methods, there are also some other approaches to dimension reduction, like principal Hessian directions (PHD)\cite{Li1992}, minimum average variance estimation(MAVE)\cite{Xia2002}, density based MAVE(dMave)\cite{Xia2007} and so on. However, in a large number of practical problems, the linear structure is often an over-ideal or approximate assumption, and the dimension reduction structure shows more nonlinear structure. So, the development of effective nonlinear dimension reduction methods is an important and meaningful research topic. The nonlinear sufficient dimension reduction replaces the linear predictor $X^{T}\bm{\beta}$ by a nonlinear predictor $U(X)$, which means model (\ref{1.1}) can be converted as follows:
\begin{equation}\label{1.2}
  F(y|X) = F(y|U(X)), \forall y \in \mathbb{R}
\end{equation}
where $U(X) = \left(u_1(X), \ldots, u_q(X)\right)^{T}, u_j \in \mathcal{H}_R(j = 1, \ldots, q)$ and $\mathcal{H}_R$ is an RKHS with reproducing kernel $R$. The complexity of the nonlinear structure increases the difficulty of research, which makes the corresponding theoretical research slow. There have been several studies on nonlinear sufficient dimension reduction methods based on kernels, such as kernel canonical component analysis (KCCA) developed by Akaho\cite{akaho2006kernel}, Bach and Jordan\cite{BachJordan2002}, Fukumizu et al\cite{fukumizu2007statistical}, and kernel dimension reduction by Fukumizu et al\cite{Kenji2009}. There have also been some methods which has developed SIR into a kernel method to deal with nonlinear sufficient dimension reduction, see \cite{HuangLee2009, Wu08, Wu2013}. While Lee \cite{Lee2013} discussed the problem in $L_2$ Space and proposed two nonlinear SDR method - GSIR and GSAVE, which do not rely on the linearity condition.

In the above methods related to model (\ref{1.1}), they all require some certain conditions. These conditions include the linearity condition, where $E(X|X^{T}\bm{\beta})$ is a linear function of X, and the constant variance condition, where $Cov(X|X^{T}\bm{\beta})$ is a constant matrix. In the methods of linear SDR, it has been shown that SIR requires the linearity condition, MAVE and dMave require $\bm{x}$ to be continues, while SAVE, PHD and DR require both linearity condition and constant variance condition. To be specific, the two conditions can be formulated as follows,
\begin{itemize}
  \item[(1)] Linearity condition : $E\left( X \big| X^{T}\bm{\beta} \right) = \mathbf{P}X$
  \item[(2)] Constant variance condition: $Cov\left( X \big| X^{T}\bm{\beta} \right) = \mathbf{Q}$
\end{itemize}
where $\mathbf{P} = \bm{\beta}(\bm{\beta}^{T}\bm{\beta})^{-1}\bm{\beta}^{T}$ and $\mathbf{Q} = \mathbf{I}_p - \mathbf{P}$.
To be free of these conditions, some new methods has been proposed. Li and Dong have successfully removed the linearity condition in the problem of dimension reduction for non-elliptically distributed predictors, see \cite{LiandDong2010,LiandDong2009}. Lee \cite{Lee2013} also proposed the methods that do not rely on the linearity condition. But they still can not get rid of the constant variance condition. Ma and Zhu\cite{MaZhu12} offered a new vision to deal with this restrict condition. By using the semiparametric theory\cite{Bickel1993,Tsiatis2006}, they have revealed that the two conditions are both unnecessary and replaced the conditions with nonparametric estimation of the corresponding conditional expectations. The semiparametric sufficient dimension reduction method proposed in \cite{MaZhu12} does not rely on the two conditions and also does not require all the covariates to be continuous. When these conditions are satisfied, but we still estimate the relevant quantities nonparametrically, Ma and Zhu \cite{ma2012efficiency} found that the resulting estimation variance of the inverse
regression method decreases and they tried to explain this puzzle in \cite{ma2013efficient,maunderstanding}.

By using the theory of RKHS, our paper extends the semiparametric method into a more generalized domain where both the interested parameters and nuisance parameters to be infinite dimensional. Motivated by the method in \cite{MaZhu12}, we derive the nuisance tangent space and its orthogonal complement under the generalized semiparametric framework. Similarly, we also construct a class of influence functions, which derive a general class of estimating equations. Ma and Zhu \cite{MaZhu12} has discussed that the application of semiparametric analysis can eliminate the linearity condition and constant variable condition. Hence, our method not only inherits these advantages but also expands the scope and applicability of linear SDR. In this paper, We take the generalized semiparametric kernel sliced inverse regression(GS-KSIR) and generalized semiparametric kernel sliced average variance estimation(GS-KSAVE) as the examples of our method. Because the dimension of interested parameters is infinity, we use the representation theorem and the regularization to make the numerical computation possible. We also establish the consistency of the estimator from the penalized estimating equation.

The outline of this article is the following. In section 2, we first describe the general semiparametric models and the approach to obtain the generalized orthogonal complement space of the nuisance tangent space and the estimating equation, then we show how to derive the GS-KSIR method. The algorithm and main theoretical results are given finally. Sections 3 and section 4 present the simulation results and the real data application. The article is finished with a brief discussion in Section 5. The proofs of main theorems are collected in an appendix.

\section{Nonlinear Sufficient Dimensional Reduction in a Generalized Semiparametric Framework \label{Sec1}}

In this section, we first study the generalized semiparametric model, whose dimension of nuisance parameters and interested parameters are all infinite. For simply, the generalized semi-parametric model can be formulated as $\{P_{\theta,\eta}: \theta \in \Theta \subset \mathcal{H}_{1}, \eta \in \mathcal{H}_{2}\}$, where $\mathcal{H}_{1}, \mathcal{H}_{2}$ are the reproducing kernel Hilbert space and we denote $\theta$, $\eta$ as the interested parameter and nuisance parameter respectively. Similar with the classical semi-parametric theory, we focus on estimating the value $\psi(P_{\theta,\eta})$ of function $\psi: \mathcal{P} \rightarrow \mathcal{H}$. Here we have $\psi(P_{\theta,\eta}) = \theta$ and we need to find a way to construct the estimation equation for the interested parameter $\theta$. Firstly, we consider the submodels $t \rightarrow P_{\theta_t, \eta_t}$ and the following definition.

\textbf{Definition 1.} A map $\psi: \mathcal{P} \rightarrow \mathcal{H}$ is differentiable at $P$ relative to a given tangent set $\dot{\mathcal{P}}_{P}$ if there exists a continuous linear map $\dot{\psi}_{P}: L_2(P) \rightarrow \mathcal{H}$ such that for every $g \in \dot{\mathcal{P}}_{P}$ and a submodel $t \rightarrow P_t$ with score function g,
\begin{equation*}
  \frac{\psi(P_t) - \psi(P)}{t} \rightarrow \dot{\psi}_{P}g
\end{equation*}
with the Riesz representation theorem for RKHS, there exists a measurable function $\tilde{\psi}_{P}: \mathcal{X} \rightarrow \mathcal{H}$ such that $\dot{\psi}_{P}g = \langle\tilde{\psi}_{P}, g \rangle_{P}$.

The score functions for the submodels $P_{\theta_t, \eta_t}$ can be expressed as
\begin{equation}\label{2.1.1}
  \frac{\partial}{\partial t}_{|t=0}\log dP_{\theta_t, \eta_t} = h + g
\end{equation}
where the function $h$ is the score function for $\theta$ when $\eta$ is fixed, $g$ is the score function for $\eta$ when $\theta$ is fixed. We denoted the tangent set for $\theta$ and $\eta$ by $\dot{\mathcal{P}}_{P_{\theta, \eta}}^{(\theta)}$ and $\dot{\mathcal{P}}_{P_{\theta, \eta}}^{(\eta)}$ respectively.

As we are interested in the parameter $\psi(P_{\theta_t, \eta_t}) = \theta_t$, then with the Definition 1 and (\ref{2.1.1}), there exists a function $\tilde{\psi}_{\theta, \eta}$ such that
\begin{equation}\label{2.1.2}
  b = \frac{\partial}{\partial t}_{|t=0}\theta_{t} = \frac{\partial}{\partial t}_{|t=0}\psi(P_{\theta_t, \eta_t}) = \langle \tilde{\psi}_{\theta, \eta}, h + g\rangle_{P_{\theta, \eta}}, \quad h \in \dot{\mathcal{P}}_{P_{\theta, \eta}}^{(\theta)}, g \in \dot{\mathcal{P}}_{P_{\theta, \eta}}^{(\eta)}
\end{equation}
When fixed $\eta$, the score functions $b$ of the submodel $t \rightarrow \theta_t$ and $h$ of the submodel $t \rightarrow P_{\theta_t, \eta}$ have the relationship as $h = A_{\theta}b$. $A_{\theta}$ is a score operator. Hence, setting $b$ in (\ref{2.1.2}) as zero, we have
\begin{equation*}
  \langle \tilde{\psi}_{\theta, \eta}, g \rangle = 0
\end{equation*}

Then we can see that the influence function $\tilde{\psi}_{\theta, \eta}$ is orthogonal to the nuisance tangent set $\dot{\mathcal{P}}_{P_{\theta, \eta}}^{(\eta)}$. This suggests that we can derive the orthogonal space of nuisance tangent space to obtain an indication of the semiparmetric estimators $T_n$ for $\theta$. Next, we will show that the problem of nonlinear sufficient dimension reduction can also be viewed as a general semiparametric model.
Suppose covariates X and response Y satisfied model (\ref{1.2}), we denote the likelihood function of one random observation (X, Y):
\begin{equation}\label{(3)}
  \eta(X,Y) = \eta_1(X)\eta_2(Y|X) = \eta_1(X)\eta_2(Y|U(X))
\end{equation}
where $\eta_1$ is a probability density function(pdf) of X, or a mixture and $\eta_2$ is the conditional pdf of Y on X. With the reproducing property of RKHS mentioned in Appendix, there exists $\bm{\beta}_j \in \mathcal{H} (j =1, \ldots, p)$ such that
\begin{equation*}
  u_j(X) = \langle u_j, R(X,\cdot)\rangle_{\mathcal{H}_R} = \langle \bm{\beta}_j, \phi(X) \rangle_{\mathcal{H}} \triangleq \bm{\beta}_j^T\phi(X), j = 1, \ldots, q.
\end{equation*}
Then we can convert $U(x)$ as follows,
\begin{equation}\label{2.2}
  U(x) = \left(\langle \bm{\beta}_1, \phi(X)\rangle_{\mathcal{H}}, \ldots, \langle \bm{\beta}_q, \phi(X)\rangle_{\mathcal{H}}\right)^{T} = \left(\bm{\beta}_1, \ldots, \bm{\beta}_q\right)^{T}\phi(X) \triangleq \bm{\beta}^T\phi(X)
\end{equation}
where $\bm{\beta} = \left(\bm{\beta}_1, \ldots, \bm{\beta}_q\right)$. Furthermore, formula (\ref{(3)}) can be simplified as
\begin{equation}\label{(4)}
  \eta_1\left(X\right)\eta_2\left(Y | \bm{\beta}^{T}\phi(X)\right)
\end{equation}
where the dimensions of interested parameter $\bm{\beta}$ and nuisance parameters $\eta_1, \eta_2$ are both infinite, which is different with the semiparametric model \cite{Bickel1993} \cite{Tsiatis2006}. Hence, nonlinear sufficient dimension reduction can be seen as a general semiparametric model.

\subsection{Nuisance Tangent Space and Its Orthogonal Complement}
To obtain the nuisance tangent space for a generalized semiparametric model, we first consider the parametric submodels,
\begin{equation*}
  \eta_1(X, \gamma_1) \text{ and } \eta_2(Y|\bm{\beta}^T\bm{\phi}(X),\gamma_2)
\end{equation*}

where $\bm{\beta} = (\beta_1, \ldots, \beta_q)$, $\beta_j \in \mathcal{H}_R$, $\gamma_1$ is an $r_1$-dimensional vector and $\gamma_2$ is an $r_2$-dimensional vector. Thus $\gamma = (\gamma_1^T, \gamma_2^T)^T$ is an r-dimensional vector, $r = r_1 + r_2$.

The parametric submodels is given as
\begin{equation}\label{(A.15)}
  \mathcal{F}_{\bm{\beta}, \gamma} = \left\{f\left(X, \bm{\beta} , \gamma_1, \gamma_2\right) = \eta_1\left(X, \gamma_1\right)\eta_2\left(Y| \bm{\beta}^T\bm{\phi}(X), \gamma_2\right)\right\} \text{ for } \left(\beta^{T}, \gamma_{1}^{T}, \gamma_{2}^{T}\right)^{T} \in \Omega_{\beta,\gamma} \subset \mathcal{H}^{q}\oplus \mathbb{R}^{r}
\end{equation}

Then the parametric submodel nuisance score vector is given as
\begin{equation*}
\begin{split}
   S_\gamma(Y,X,\bm{\beta}_0,\gamma_0) & = \left\{ \left(\frac{\partial f(X,Y,\bm{\beta},\gamma)}{\partial \gamma_1}\right)^T, \left(\frac{\partial f(X,Y,\bm{\beta},\gamma)}{\partial \gamma_2}\right)^T \right\}^T \Bigg|_{\bm{\beta} = \bm{\beta}_0, \gamma = \gamma_0} \\
   ~ & = \left\{S_{\gamma_1}^{T}(X,Y,\bm{\beta}_0,\gamma_0), S_{\gamma_2}^{T}(X,Y,\bm{\beta}_0,\gamma_0) \right\}^T
\end{split}
\end{equation*}
where
\begin{equation*}
  S_{\gamma_1}^{T}(X,Y,\bm{\beta}_0,\gamma_0) = \frac{\partial \log f_X(X, \gamma_1)}{\partial \gamma_1} \quad \text{ and } \quad S_{\gamma_2}^{T}(X,Y,\bm{\beta}_0,\gamma_0) = \frac{\partial \log f_{Y|X}(Y,\bm{\beta}^T\bm{\phi}(X),\gamma_2)}{\partial \gamma_2}
\end{equation*}

The elements in parametric submodel nuisance tangent space can be given by
\begin{equation*}
  B^{q \times r} S_{\gamma}(X,Y) = B_1^{q \times r_1} S_{\gamma_1}(X) + B^{q \times r_2} S_{\gamma_2}(X,Y)
\end{equation*}
Hence, the parametric submodel nuisance tangent space can be written as $\Lambda_\gamma = \Lambda_{\gamma_1} \oplus \Lambda_{\gamma_2}$, where $\Lambda_{\gamma_1}$ and $\Lambda_{\gamma_2}$ are the parametric submodel nuisance tangent space corresponding to $\eta_1$ and $\eta_2$, which are given as
\begin{equation*}
  \Lambda_{\gamma_1} = \{B^{q \times r_1}S_{\gamma_1}(X) \text{ for all } B^{q \times r_1}\}
\end{equation*}
and
\begin{equation*}
  \Lambda_{\gamma_2} = \{B^{q \times r_2}S_{\gamma_2}(X,Y) \text{ for all } B^{q \times r_1}\}
\end{equation*}

From the theory of semiparametric methods \cite{Bickel1993} \cite{Tsiatis2006}, we have the definition of the nuisance tangent space $\Lambda$, which is a subspace of $\mathcal{H}$ and defined as the mean squared closure of all the elements of the form $\left\{B^{q \times r}S_{\gamma}^{r \times 1}(X,\bm{\beta}_0, \gamma_0)\right\}$, i.e.
\begin{equation*}
  \begin{split}
     \Lambda = & \left\{h^{q \times 1}(X)\in \mathcal{H} \text{ such that } E\left\{h^{T}(X)h(X)\right\} < \infty \text{ and there exists }\right. \\
       & \left.\text{a sequence } B_{j}S_{\gamma j}(X) \text{ such that } \|h(X) - B_{j}S_{\gamma j}(X)\| \longrightarrow 0 \text{ as } j \rightarrow \infty\right\}
  \end{split}
\end{equation*}
where S is an arbitrary nuisance score vector function, B is any conformable matrix with q rows and $\|h(X)\|^2 = E\left\{h^{T}(X)h(X)\right\}$. Same with the definition of $\Lambda$, we can derive the tangent space corresponding to $\eta_1$ and $\eta_2$ as follows.
\begin{lemma}\label{lemma2.1}
  The nuisance tangent space corresponding to $\eta_1$ and $\eta_2$ is $\Lambda_1 = \{\text{mean-square closure of all } \Lambda_{\gamma_1}\}$ consists of all q-dimensional mean-zero functions of X, i.e.
  \begin{equation*}
    \Lambda_1 = \left\{f: E(f) = 0\right\}
  \end{equation*}
  and $\Lambda_2 = \{\text{mean-square closure of all } \Lambda_{\gamma_2}\}$ is the space of all q-dimensional random functions $f(Y,X)$ that satisfy $E(f|X) = E(f|U(X)) = 0$, i.e.
  \begin{equation*}
  \Lambda_2 = \left\{f: E(f|X)=E(f|U(X))= 0 \right\}
\end{equation*}
\end{lemma}

Lemma \ref{lemma2.1} gives us the nuisance tangent space of $\eta_1$ and $\eta_2$. Next, we derive the their orthogonal complement $\Lambda_{1}^{\perp}$ and $\Lambda_{2}^{\perp}$. With the orthogonal complement $\Lambda_{1}^{\perp}$ and $\Lambda_{2}^{\perp}$, we can obtain the generalized nuisance tangent space orthogonal complement of model (\ref{1.2}). The detail is given in the following Theorem \ref{thm1}.

\begin{theorem}\label{thm1}
  For the generalized semiparamtric model (\ref{1.2}), we denote by $\Lambda^{\perp}$ the generalized nuisance tangent space orthogonal complement (GNTSOC for shortly) of this model, then
  \begin{equation*}
    \Lambda^{\perp} = \left\{ f(Y,X) - E(f|U(X),Y) ~\Big|~ E(f|X) = E(f|U(X)), \forall f\right\}
  \end{equation*}
  where , $U(X) = \left(u_1(X), \ldots, u_q(X)\right)^{T}, u_j \in \mathcal{H}_R(j = 1, \ldots, q)$ and $\mathcal{H}_R$ is an RKHS with reproducing kernel $R$.
\end{theorem}

Theorem \ref{thm1} describes the form of the elements in GNTSOC, then we can construct the $f(Y,X)$ which satisfies $E(f|X) = E(f|U(X))$ and such $f(Y,X)$ can be used to construct the estimation equation of interested parameters. The detail derivation of Lemma \ref{lemma2.1} and Theorem \ref{thm1} is given in Appendix.

\subsection{Estimation Equation of Nonlinear Sufficient Dimensional Reduction}
In Section 2.1, we have discussed the nuisance tangent space and its orthogonal complement, which gives us the many possible ways to consistent estimating equations. Then in this section, we will show some ways to construct estimation equations from the GNTSOC. Similar with the construct ways in \cite{MaZhu12}, function $f(Y,X)$ in $\Lambda^{\perp}$ can be constructed as follows: for any function $g\left(Y,U(X)\right)$ and $a(X)$,
\begin{equation}\label{(5)}
  \left[g(Y, U(X)) - E\left\{g(Y, U(X))|U(X)\right\}\right] \times \left[a(X) - E\left(a(X)|U(X)\right)\right]
\end{equation}
It's easy to see that such $f(X,Y)$ satisfies $E(f|U(X),Y) = 0$ and $E(f|X) = E(f|U(X)) = 0$, then $f(Y,X)$ is an element in $\Lambda^{\perp}$.
With the equation (\ref{2.2}), (\ref{(5)}) can be converted as
\begin{equation}\label{(6)}
  \left[g(Y, \bm{\beta}^{T}\phi(X)) - E\left\{g(Y, \bm{\beta}^{T}\phi(X))|\bm{\beta}^{T}\phi(X))\right\}\right] \times \left[a(X) - E\left(a(X)|\bm{\beta}^{T}\phi(X)\right)\right]
\end{equation}
Then we can construct the following equation
\begin{equation}\label{(7)}
  E\left\{ \left[g(Y, \bm{\beta}^{T}\phi(X)) - E\left\{g(Y, \bm{\beta}^{T}\phi(X))|\bm{\beta}^{T}\phi(X))\right\}\right] \otimes \left[a(X) - E\left(a(X)|\bm{\beta}^{T}\phi(X)\right)\right] \right\} = 0
\end{equation}

Next we choose generalized semiparametric kernel sliced inversed regression (GS-KSIR for shortly) as a special case of our method.

To obtain the estimation equation of GS-KSIR, we set that $g(Y, \bm{\beta}^{T}\phi(X)) = E(\phi(X)|Y)$ and $a(X) = \phi^{T}(X)$. Then we have
\begin{equation}\label{(8)}
  E\left\{ \left[E(\phi(X)|Y) - E\left\{E(\phi(X)|Y)|\bm{\beta}^{T}\phi(X))\right\}\right] \otimes \left[\phi(X) - E\left(\phi(X)|\bm{\beta}^{T}\phi(X)\right)\right] \right\} = 0
\end{equation}
To simplify (\ref{(8)}), we set
\begin{equation*}
  E(\phi(X)|Y) = \xi(Y), \quad E\left\{E(\phi(X)|Y)|\bm{\beta}^{T}\phi(X))\right\} = \zeta(\bm{\beta}^{T}\phi(X)), \quad E\left(\phi(X)|\bm{\beta}^{T}\phi(X)\right) = \theta(\bm{\beta}^{T}\phi(X))
\end{equation*}

and their corresponded kernel estimation to be $\hat{\xi}(Y)$, $\hat{\zeta}(\bm{\beta}^{T}\phi(X))$ , $\hat{\theta}(\bm{\beta}^{T}\phi(X))$.
When given the observations $\{(x_i, y_i)\}_{i=1}^{n}$, we can obtain the sample version of equation (\ref{(8)}) as
\begin{equation}\label{(9)}
  \frac{1}{n}\sum_{i=1}^{n}\left[\left(\hat{\xi}(y_i) - \hat{\zeta}(\bm{\beta}^{T}\phi(x_i))\right) \times \left(\phi(x_i) - \hat{\theta}(\bm{\beta}^{T}\phi(x_i))\right)\right] = 0
\end{equation}
where
\begin{equation*}
  \hat{\xi}(y_i) = \frac{\sum_{j=1}^{n}\phi(x_j)K_{h_1}^{(1)}(y_j, y_i)}{\sum_{j=1}^{n}K_{h_1}^{(1)}(y_j, y_i)}
\end{equation*}

\begin{equation*}
  \hat{\zeta}(\beta^T\phi(x_i)) = \sum_{j=1}^{n}\hat{\xi}(y_i)\frac{K_{h_2}^{(2)}(\beta^T\phi(x_j), \beta^T\phi(x_i))}{\sum_{j=1}^{n}K_{h_2}^{(2)}(\beta^T\phi(x_j), \beta^T\phi(x_i))}
\end{equation*}

\begin{equation*}
  \hat{\theta}(\beta^T\phi(x_i)) = \sum_{j=1}^{n}\phi(x_j)\frac{K_{h_2}^{(2)}(\beta^T\phi(x_j), \beta^T\phi(x_i))}{\sum_{j=1}^{n}K_{h_2}^{(2)}(\beta^T\phi(x_j), \beta^T\phi(x_i))}
\end{equation*}
We can convert equation (\ref{(9)}) to
\begin{equation}\label{(10)}
  \bm{\Phi} \tilde{\mathbf{K}}_{h_1}^{(1)}(\mathbf{I}_n - \tilde{\mathbf{K}}_{h_2}^{(2)})(\mathbf{I}_n - \tilde{\mathbf{K}}_{h_2}^{(2)})^{T}\bm{\Phi}^{T} = 0
\end{equation}
where
\[ \bm{\Phi} = (\phi(x_1)), \ldots, \phi(x_n)), \]

\[ \tilde{\mathbf{K}}_{h_1}^{(1)} = \left[ \frac{K_{h_1}^{(1)}(y_i, y_j)}{\sum_{\ell = 1}^{n}K_{h_1}^{(1)}(y_\ell, y_j)}\right]_{1 \le i,j \le n},  \quad
\tilde{\mathbf{K}}_{h_2}^{(2)} = \left[\frac{K_{h_2}^{(2)}(\bm{\beta}^{T}\phi(x_i), \bm{\beta}^{T}\phi(x_j))}{\sum_{\ell = 1}^{n}K_{h_2}^{(2)}(\bm{\beta}^{T}\phi(x_\ell), \bm{\beta}^{T}\phi(x_j))}\right]_{1 \le i,j \le n}\]
where $h_1, K_{h_1}^{(1)}$ are the bandwidth and kernel function in the nonparametric estimation of $E(\phi(X)|Y)$, respectively; $h_2, K_{h_2}^{(2)}$ are the bandwidth and kernel function in the nonparametric estimation of $E(\phi(X)|\bm{\beta}^{T}\phi(X))$, respectively.

Before solving the estimation equation (\ref{(10)}), we first centralize the mapped data $\bm{\Phi}$ by $\tilde{\bm{\Phi}} = \bm{\Phi}(\bm{I}_{n} - \frac{1}{n}\bm{1}_{n}\bm{1}_{n}^{T})$. Then we have
\begin{equation}\label{10_1}
  \tilde{\bm{\Phi}} \tilde{\mathbf{K}}_{h_1}^{(1)}(\mathbf{I}_n - \tilde{\mathbf{K}}_{h_2}^{(2)})(\mathbf{I}_n - \tilde{\mathbf{K}}_{h_2}^{(2)})^{T}\tilde{\bm{\Phi}}^{T} = 0
\end{equation}
Because the dimension of $\phi(X)$ is infinite, it is not possible to solve equation (\ref{10_1}) and obtain the estimator $\hat{\bm{\beta}}$ directly. Firstly, we try to deal with the quantity $\bm{\beta}^{T}\phi(x)$.

For $\forall j = 1, \ldots, q$, we have $\beta_j = \beta_j^{(0)} + \beta_j^{(1)}$, where $\beta_j^{(0)} \in \mathcal{H}_0 = \overline{span}\{\phi(x_1), \ldots, \phi(x_n)\}$ and $\beta_j^{(1)} \in \mathcal{H}_{0}^{\perp}$. Here, $\mathcal{H} = \mathcal{H}_0 \oplus \mathcal{H}_{0}^{\perp}$.

Then, there exists $c_j \in \mathbb{R}^n$ such that
\begin{equation*}
  \beta_{j}^{(0)} = \sum_{i=1}^{n}c_{ij}\phi(x_i) = \mathbf{\Phi}c_j
\end{equation*}
So, we have
\begin{align*}
  \langle \beta_j, \phi(x_i)\rangle_{\mathcal{H}} & = \langle \beta_j^{(0)} + \beta_j^{(1)}, \phi(x_i)\rangle_{\mathcal{H}} = \langle \beta_j^{(0)}, \phi(x_i)\rangle_{\mathcal{H}}\\
  ~ & = \sum_{\ell = 1}^{n}c_{\ell j}\langle\phi(x_\ell), \phi(x_i)\rangle_{\mathcal{H}} = \sum_{\ell=1}^{n}c_{\ell j}R(x_\ell, x_i)\\
  ~ & = \left(R(x_1, x_i), \ldots, R(x_n, x_i)\right)^{T}c_j
\end{align*}
We denote the $i$th column of Gram matrix $(R(x_i, x_j))_{1 \le i,j \le n}$ as $R_i$, then
\begin{equation*}
  \beta^T\mathbf{\Phi}(x_i) = \begin{pmatrix}
                                \langle \beta_1, \phi(x_i)\rangle_{\mathcal{H}} \\
                                \vdots \\
                                \langle \beta_q, \phi(x_i)\rangle_{\mathcal{H}}
                              \end{pmatrix}
                            = \begin{pmatrix}
                                c_{1}^{T} \\
                                \vdots \\
                                c_{q}^{T}
                              \end{pmatrix}R_{(i)}
                            = \bm{C}^{T}R_{(i)}
\end{equation*}
where $\bm{C} = (\bm{C}_1, \ldots, \bm{C}_q), \bm{C}_j \in \mathbb{R}^{n}(j = 1,\ldots,q)$; $R_{(i)}$ is the $i$th column of Gram matrix $(K(x_i, x_j))_{1 \le i,j \le n}$. Then the $\tilde{\mathbf{K}}_{h_2}^{(2)}$ in (\ref{10_1}) can be converted as
\begin{equation}\label{(11)}
  \tilde{\mathbf{K}}_{h_2}^{(2)} = \left[ \frac{K_{h_2}^{(2)}(\bm{C}^{T}R_{(i)}, \bm{C}^{T}R_{(j)})}{\sum_{\ell = 1}^{n}K_{h_2}^{(2)}(\bm{C}^{T}R_{(\ell)}, \bm{C}^{T}R_{(j)})} \right]_{1 \le i,j \le n}.
\end{equation}

After dealing with the $\bm{\beta}^{T}\phi(x)$, we consider the following theorem.
\begin{theorem}\label{thm2}
  Equation
  \begin{equation*}
      \tilde{\bm{\Phi}} \tilde{\mathbf{K}}_{h_1}^{(1)}(\mathbf{I}_n - \tilde{\mathbf{K}}_{h_2}^{(2)})(\mathbf{I}_n - \tilde{\mathbf{K}}_{h_2}^{(2)})^{T}\tilde{\bm{\Phi}}^{T} = 0
  \end{equation*}
  and
  \begin{equation}\label{(12)}
    \tilde{\mathbf{R}}\tilde{\mathbf{K}}_{h_1}^{(1)}(\mathbf{I}_n - \tilde{\mathbf{K}}_{h_2}^{(2)})(\mathbf{I}_n - \tilde{\mathbf{K}}_{h_2}^{(2)})^{T}\tilde{\mathbf{R}} = 0
  \end{equation}
  are equivalent, where $\tilde{\mathbf{R}} = \tilde{\bm{\Phi}} ^{T}\tilde{\bm{\Phi}}$.
\end{theorem}

Combining Theorem \ref{thm2} and equation (\ref{(11)}), the problem of generalized semiparametric kernel sliced inversed regression is converted to solve the equation (\ref{(12)}). Hence we have already found a way to obtain the GS-KSIR method. While Li and Dong \cite{LiandDong2009} modified the classical SIR and proposed a new SIR for non-elliptical predictors, which does not rely on linearity condition. Their results also offers us another perspective to see the GS-KSIR and we can also extend their method to obtain the modified GS-KSIR. Li and Dong \cite{LiandDong2009} tried to recover the SDR subspace $S_{Y|x}$ through minimizing
\begin{equation*}
  E\left(\left\|E(x|Y) - E\left\{E\left(x|x^{T}\bm{\beta}\right)|Y\right\}\right\|^2\right)
\end{equation*}
Then the above equality is equivalent to
\begin{align*}
  E\left(\left[E(x|Y) - E\{E(x|x^{T}\bm{\beta})|Y\}\right]^{T}E\left\{\frac{\partial E(x|x^{T}\bm{\beta})}{\partial {vec(\bm{\beta})}^{T}}|Y\right\}\right) & = 0\\
  E\left(\left\{x - E(x|x^{T}\bm{\beta})\right\}^{T}E\left\{\frac{\partial E(x|x^{T}\bm{\beta})}{\partial {vec(\bm{\beta})}^{T}}|Y\right\}\right) & = 0
\end{align*}
Next, we consider the nonlinear SDR subspace $S_{Y|U(x)}$ and try to recover $S_{Y|U(x)}$ through minimizing
\begin{equation}\label{13}
  E\left(\left\|E(\phi(x)|Y) - E\left\{E\left(\phi(x)|\bm{\beta}^{T}\phi(x)\right)|Y\right\}\right\|^2\right)
\end{equation}
The above minimization is equivalent to
\begin{equation*}
  E\left(\left[E(\phi(x)|Y) - E\left\{E(\phi(x)|\bm{\beta}^{T}\phi(x))|Y\right\}\right]\otimes E\left\{\frac{\delta E(\phi(x)|\bm{\beta}^{T}\phi(x))}{\delta \bm{\beta}}|Y\right\}\right)  = 0
\end{equation*}
With the law of iterated expectation, we have
\begin{equation*}
  E\left(\left[\phi(x) - E(\phi(x)|\bm{\beta}^{T}\phi(x))\right]\otimes E\left\{\frac{\delta E(\phi(x)|\bm{\beta}^{T}\phi(x))}{\delta \bm{\beta}}|Y\right\}\right)  = 0
\end{equation*}
where $\delta E(\phi(x)|\bm{\beta}^{T}\phi(x))/\delta \bm{\beta}$ is the Fr\'{e}chet derivative of $E(\phi(x)|\bm{\beta}^{T}\phi(x))$. To see the above method as a special case of the (\ref{(7)}), we can choose $a(x) = \phi(x) - E(\phi(x)|\bm{\beta}^{T}\phi(x))$ and $g(Y, \bm{\beta}^{T}\phi(x)) = E\left\{\delta E(\phi(x)|\bm{\beta}^{T}\phi(x))/\delta \bm{\beta}\big|Y\right\}$.

To be convenient, we set
$\xi(Y) = E(\phi(x)|Y)$, $\theta(\beta^T\phi(x)) = E(\phi(x)|\bm{\beta}^{T}\phi(x) \rangle)$, $\psi(Y) = E\left\{E(\phi(x)|\bm{\beta}^{T}\phi(x))|Y\right\}$ and their corresponded kernel estimation to be $\widehat{\xi}(Y)$, $\widehat{\theta}(\beta^T\phi(x))$, $\widehat{\psi}(Y)$. Similar with the derivation of (\ref{(9)}) and (\ref{(10)}), we obtain the sample version of (\ref{13}):
\begin{equation}\label{14}
  \min_{\bm{C}}\frac{1}{n}\left\|\hat{\xi}(y_i) - \hat{\psi}(y_i)\right\|^2
\end{equation}
where
\begin{equation*}
  \hat{\xi}(y_i) = \frac{\sum_{j=1}^{n}\phi(x_j)K_{h_1}^{(1)}(y_j, y_i)}{\sum_{j=1}^{n}K_{h_1}^{(1)}(y_j, y_i)}
\end{equation*}

\begin{equation*}
  \hat{\theta}(\beta^T\phi(x_i)) = \sum_{j=1}^{n}\phi(x_j)\frac{K_{h_2}^{(2)}(\beta^T\phi(x_j), \beta^T\phi(x_i))}{\sum_{j=1}^{n}K_{h_2}^{(2)}(\beta^T\phi(x_j), \beta^T\phi(x_i))} = \sum_{j=1}^{n}\phi(x_j)\frac{K_{h_2}^{(2)}(\bm{C}^{T}R_{(j)}, \bm{C}^{T}R_{(i)})}{\sum_{j=1}^{n}K_{h_2}^{(2)}(\bm{C}^{T}R_{(j)}, \bm{C}^{T}R_{(i)})}
\end{equation*}

\begin{equation*}
  \hat{\psi}(y_i) = \sum_{j=1}^{n}\hat{\theta}(\beta^T\phi(x_i))\frac{K_{h_3}^{(1)}(y_j, y_i)}{\sum_{j=1}^{n}K_{h_3}^{(1)}(y_j, y_i)}
\end{equation*}
minimization (\ref{14}) can be converted to
\begin{equation}\label{15}
  \min_{\bm{C}}\bm{\Phi}^{T}\left(\tilde{\mathbf{K}}_{h_1}^{(1)} - \tilde{\mathbf{K}}_{h_2}^{(2)}\tilde{\mathbf{K}}_{h_3}^{(1)}\right)\left(\tilde{\mathbf{K}}_{h_1}^{(1)} - \tilde{\mathbf{K}}_{h_2}^{(2)}\tilde{\mathbf{K}}_{h_3}^{(1)}\right)^{T}\bm{\Phi}
\end{equation}
which is equivalent to
\begin{equation}\label{16}
  \min_{\bm{C}}\mathbf{R}\left(\tilde{\mathbf{K}}_{h_1}^{(1)} - \tilde{\mathbf{K}}_{h_2}^{(2)}\tilde{\mathbf{K}}_{h_3}^{(1)}\right)\left(\tilde{\mathbf{K}}_{h_1}^{(1)} - \tilde{\mathbf{K}}_{h_2}^{(2)}\tilde{\mathbf{K}}_{h_3}^{(1)}\right)^{T}\mathbf{R}
\end{equation}

We have presented two ways to derive the GS-KSIR, both of them can be seen as the special cases of the generalized semiparametric method (\ref{(7)}). Furthermore, we can also derive other nonlinear SDR method, like generalized semiparametric kernel sliced average
variance estimation (GS-KSAVE for shortly). The detail of GS-KSAVE are presented in the Appendix.

\subsection{Theorem and Algorithm}

\textbf{Notation:} Given i.i.d. random variables $X_{1}, \ldots, X_{n}$ with law $P$ on a measurable space
$(\mathcal{X}, \mathcal{A})$ and a measurable function $f:\mathcal{X} \rightarrow \mathbb{R}$  we let $\mathbb{P}_{n}f = \frac{1}{n}\sum_{i=1}^{n}f(X_{i})$, $Pf = \int f dP$ and $\mathbb{G}_{n}f = \sqrt{n}\left(\mathbb{P}_{n} - P\right)f$.

As the dimension of $\bm{\beta}$ is infinity, we consider the penalized estimation equation and establish the consistency of $\widehat{\bm{\beta}}$ obtained from the penalized estimation equation in the following Theorem \ref{thm3}.
\begin{theorem}\label{thm3}
  Under conditions (C1)-(C4) given in Appendix, with the estimator $\widehat{\bm{\beta}}$ obtained from the estimation equation
  \begin{equation}\label{thm3_eq1}
    \sum_{i=1}^{n}\left[g(Y_i, \widehat{\bm{\beta}}^{T}\phi(x_i)) - \widehat{E}\{g(Y_i, \widehat{\bm{\beta}}^{T}\phi(x_i))|\widehat{\bm{\beta}}^{T}\phi(x_i)\}\right]\left[a(x_i) - \widehat{E}\{a(x_i)|\widehat{\bm{\beta}}^{T}\phi(x_i)\}\right] + n\widehat{\bm{\beta}}\bm{\Gamma} = 0
  \end{equation}
  we have
  \begin{equation*}
    \left\|\widehat{\bm{\beta}} - \bm{\beta}\right\| \stackrel{Pr}\longrightarrow 0
  \end{equation*}
  where $\bm{\beta}$ is the true value and $\bm{\Gamma} = diag\{\gamma_1, \ldots, \gamma_q\}$
\end{theorem}

Theorem \ref{thm3} ensures us to obtain the consistency estimator from penalized estimation equation and its detail proof is given in the appendix A.6. To obtain the asymptotic distribution of $\bm{\beta}$, we refer to the following Theorem.
\begin{theorem}\label{thm4}
Suppose that the class of functions $\{\psi_{\theta,\eta}: \theta \in \Theta, \eta \in \mathcal{H}\}$ is P -Donsker, that the map $\theta \rightarrow P\psi_{\theta}$ is $Fr\acute{e}chet$ differentiable at $\theta_0$ with
derivative $A: lin\Theta \rightarrow \ell^{\infty}(\mathcal{H})$. Furthermore, assume that the maps $\theta \rightarrow \psi_{\theta, \eta}$ are continuous in $\ell_{2}(P)$ at $\theta_0$, uniformly in $\eta \in \mathcal{H}$. If $\lim_{n \rightarrow \infty}\sqrt{n}\gamma_{n} = 0$ and penalty function $q(\cdot)$ is bounded, Then any zero $\hat{\theta}_{n}$ of $\theta \rightarrow \mathbb{P}_{n}\psi_{\theta} + \gamma_{n}q(\theta)$ that converges in probability to a zero $\theta_{0}$ of $\theta \rightarrow P\psi_{\theta}$ satisfies
\begin{equation*}
  A\sqrt{n}(\hat{\theta}_{n} - \theta_{0}) = \mathbb{G}_{n}\psi_{\theta_{0}} + o_p(1)
\end{equation*}
\end{theorem}
We define
\begin{align*}
  P\psi_{\bm{\beta}} & = E\left\{ \left[g(Y, \bm{\beta}^{T}\phi(X)) - E\left\{g(Y, \bm{\beta}^{T}\phi(X))|\bm{\beta}^{T}\phi(X))\right\}\right] \left[a(X) - E\left(a(X)|\bm{\beta}^{T}\phi(X)\right)\right] \right\} \\
  \mathbb{P}_{n}\tilde{\psi}_{\widehat{\bm{\beta}}} & = \frac{1}{n}\sum_{i=1}^{n}\left[g(Y_i, \widehat{\bm{\beta}}^{T}\phi(x_i)) - \widehat{E}\{g(Y_i, \widehat{\bm{\beta}}^{T}\phi(x_i))|\widehat{\bm{\beta}}^{T}\phi(x_i)\}\right]\left[a(x_i) - \widehat{E}\{a(x_i)|\widehat{\bm{\beta}}^{T}\phi(x_i)\}\right] \\
  \mathbb{P}_{n}\psi_{\widehat{\bm{\beta}}} & = \frac{1}{n}\sum_{i=1}^{n}\left[g(Y_i, \widehat{\bm{\beta}}^{T}\phi(x_i)) - E\{g(Y_i, \widehat{\bm{\beta}}^{T}\phi(x_i))|\widehat{\bm{\beta}}^{T}\phi(x_i)\}\right]\left[a(x_i) - E\{a(x_i)|\widehat{\bm{\beta}}^{T}\phi(x_i)\}\right]
\end{align*}
in Theorem \ref{thm4} and we obtain $\mathbb{P}_{n}\tilde{\psi}_{\widehat{\bm{\beta}}} - \mathbb{P}_{n}\psi_{\widehat{\bm{\beta}}} = o_p(1)$ from the derivation in Theorem \ref{thm3}. Then estimator $\widehat{\bm{\beta}}$ from (\ref{thm3_eq1}) has the asymptotic distribution
\begin{equation*}
  \sqrt{n}A(\widehat{\bm{\beta}} - \bm{\beta}) \rightsquigarrow \mathbb{G}\psi_{\bm{\beta}}
\end{equation*}

Next, we will show the algorithm to calculate estimator numerically. To obtain the solution of equation (\ref{(12)}) and make the algorithm achieve stability, we refer to the regular methods similar to that in \cite{Li2008, Zhong2005}. The objective function is
\begin{equation}\label{algorithm01}
  Q(\mathbf{C}) \triangleq \left\|\tilde{\mathbf{R}}\tilde{\mathbf{K}}_{h_1}^{(1)}\left(\mathbf{I}_n - \tilde{\mathbf{K}}_{h_2}^{(2)}\right)\left(\mathbf{I}_n - (\tilde{\mathbf{K}}_{h_2}^{(2)})^{T}\right)\tilde{\mathbf{R}}\right\|^2 + \lambda \cdot tr\left(\mathbf{C}^{T}\mathbf{C}\right)
\end{equation}
Similarly, to obtain the solution for modified GS-KSIR, we centralize the matrix $\mathbf{R}$ and add the regular condition to (\ref{16}). Then the objective function can be formulated as follows.
\begin{equation}\label{algorithm02}
  Q(\mathbf{C}) \triangleq \left\|\tilde{\mathbf{R}}\left(\tilde{\mathbf{K}}_{h_1}^{(1)} - \tilde{\mathbf{K}}_{h_2}^{(2)}\tilde{\mathbf{K}}_{h_3}^{(1)}\right)\left(\tilde{\mathbf{K}}_{h_1}^{(1)} - \tilde{\mathbf{K}}_{h_2}^{(2)}\tilde{\mathbf{K}}_{h_3}^{(1)}\right)^{T}\tilde{\mathbf{R}}\right\|^2 + \lambda \cdot tr\left(\mathbf{C}^{T}\mathbf{C}\right)
\end{equation}

Then we minimize the objective function $Q(\mathbf{C})$ and obtain $\widehat{\mathbf{C}}$. The steps of algorithm is given in the following part.

\begin{itemize}
  \item[(1)] Calculate Gram matrix $\mathbf{R} = \left(R(x_i, x_j)\right)_{1 \le i,j \le n}$ and centralize $\mathbf{R}$ by $(\bm{I}_{n} - \frac{1}{n}\bm{1}_{n}\bm{1}_{n}^{T})\mathbf{R}(\bm{I}_{n} - \frac{1}{n}\bm{1}_{n}\bm{1}_{n}^{T})$.
  \item[(2)] Use the cross-validation method to select a suitable $\lambda$:

      Split the data $(X_i, Y_i),i=1,\ldots,n$ into k folds, denote $(X^{(\ell)}, Y^{(\ell)})$ and $(X^{(-\ell)}, Y^{(-\ell)})$ as the $\ell$th part and the whole data with $(X^{(\ell)}, Y^{(\ell)})$ removed respectively;

      \textbf{For} $\ell=1:k$
        \begin{itemize}
          \item[1.]Calculate $\widehat{\bm{C}}^{(-\ell)}$ based on data $(X^{(-\ell)}, Y^{(-\ell)})$ and $r_{\ell} = KCCA\left(\widehat{\bm{C}}^{(-\ell)^T}R(X^{(-\ell)},X^{(\ell)}), Y^{(\ell)}\right)$, where $R(X^{(-\ell)},X^{(\ell)}) = R(x_i,x_j), x_i \in X^{(-\ell)}, x_j \in X^{(\ell)}$;
        \end{itemize}
      \textbf{end};\\
      Maximize the $CV(\lambda) = \frac{1}{k}\sum_{\ell=1}^{k}r_{\ell}$.
  \item[(3)] While given $\lambda$, initialize $\widehat{\mathbf{C}}$ and use Newton-Rapson algorithm to obtain the solution of the minimization $\min_{\bm{C}} Q(\bm{C})$. Here we choose the solution of KSIR to be the initialization value of GS-KSIR.
\end{itemize}

As lacking of the true direction of dimension reduction in the cross-validation procedure, we use Kernel Canonical Correlation Analysis(KCCA) to evaluate the correlation between $y$ and $\widehat{\bm{U}}(x)$, which can be regarded as the replacement of the correlation between $\bm{U}(x)$ and $\widehat{\bm{U}}(x)$. The cross-validation criterion can also be formulated as $CV(\lambda) = \frac{1}{n}\sum_{\ell=1}^{n}\|y_{\ell} - \hat{f}(\widehat{\bm{C}}^{(-\ell)^T}R(X^{(-\ell)},X^{(\ell)}))\|^2$.
Then the cross-validation strategy is to predict $f(\widehat{\bm{C}}^{(-\ell)^T}R(X^{(-\ell)},X^{(\ell)}))$ for each $\ell$, see \cite{Lee2013}.

\section{Simulation Studies \label{Sec3}}

In this section, we conduct several simulation studies to evaluate the performance of different methods. we consider three cases to show our method in different situations, where the covariates X do or do not satisfy both the linearity condition and constant variance condition, and the structure of dimension reduction is linear or nonlinear. In each case, we repeat our simulations $N=100$ times with samples size $n=200$ and covariate dimension $p = 10, 20$ . To choose bandwidth for kernel estimation, We refer to \cite{MaZhu12} and use the Epanechnikov kernel and the default bandwidth selector implemented in Matlab routine ksdensity. To show the advantages of our method, we compare KSIR, S-SIR with GS-KSIR-I and GS-KSIR-II, where GS-KSIR-I and GS-KSIR-II is derived from (\ref{algorithm01}) and (\ref{algorithm02}) respectively. Moreover, We refer to the correlation $\bar{r}^2$ described in \cite{LiandDong2009} to evaluate estimation accuracy of dimension reduction, which is formulated as follows:
\begin{equation*}
  \bar{r}^2 = \frac{1}{N}\sum_{i=1}^{N}r^2\left(\bf{U}(x),\widehat{\bf{U}}(x)\right)
\end{equation*}
where
\begin{equation*}
  r^2(\bf{u},\bf{v}) = k^{-1}\sum_{i=1}^{k}\lambda_i
\end{equation*}
and $\lambda_1, \ldots, \lambda_k$ is the nonzero eigenvalues of
\begin{equation*}
  \left\{var(\bf{u})\right\}^{-1/2}cov\left(\bf{u},\bf{v}^{T}\right)\left\{var(\bf{v})\right\}^{-1}cov\left(\bf{v},\bf{u}^{T}\right)\left\{var(\bf{u})\right\}^{-1/2}
\end{equation*}

Next, we will show the three cases and their results.

Case 1: We consider the situation that the dimension reduction structure is nonlinear and covariate X violates linearity condition and constant variance condition. The data is generated from the following process:
\begin{equation*}
  Y = \left|\sum_{i=1}^{10}X_i^2\right| + 2 \times exp\left(\sum_{i=1}^{10}(-1)^{i+1}X_i^2\right) + 0.1\epsilon
\end{equation*}

where covariate X is generated by
\begin{align*}
  (X_1,X_2,X_3,X_4) & \sim N(0,\Sigma), \Sigma = (\sigma_{i,j})_{4 \times 4}, \sigma_{i,j} = 0.5^{|i-j|}\\
   X_5 & = |X_1 + X_2| + |X_1|\epsilon_1 \\
   X_6 & = |X_1 + X_2|^2 + |X_2|\epsilon_2 \\
   X_7 & \sim B(1,exp(X_2)/\left(1+exp(X_2))\right)\\
   X_8 & \sim B(1,\Phi(X_2)), \Phi(\cdot) \text{ is the CDF of standard normal distribution}\\
   X_9 & = X_3^3 - 2|X_4| + |X_3|\epsilon_3\\
   X_{10} & = |X_3+X_4|^2 + |X_4|\epsilon_4\\
   \epsilon_1,\epsilon_2,\epsilon_3,\epsilon_4,\epsilon & \sim \text{ i.i.d } N(0,1)\\
   (X_{11},\ldots,X_{p}) & \sim N(0,\Sigma), \Sigma = (\sigma_{i,j})_{(p-10) \times (p-10)}, \sigma_{i,j} = 0.6^{|i-j|}
\end{align*}

Case 2: We consider the situation that the dimension reduction structure is nonlinear and covariate X satisfies linearity condition and constant variance condition. In this case covariate X is generated from multivariate Gaussian distribution with zero mean and covariance structure such that $cov(X_i,X_j) = \sigma_{i,j} = 0.8^{|i-j|}, 1\le i,j \le p$. The model is
\begin{equation*}
  Y = \left|\sum_{i=1}^{10}X_i^2\right|\times\left(\sum_{i=1}^{10}(-1)^{i+1}X_i^2\right) + 0.5\epsilon
\end{equation*}

Case 3: We consider the situation that the dimension reduction structure is linear and covariate X violates linearity condition and constant variance condition. In this case, covariate X is same with that in case 1 and Y is generated from
\begin{equation*}
  Y = \left(\sum_{i=1}^{10}X_i\right)^{2} + \left(\sum_{i=1}^{10}(-1)^{i+1}X_i\right)^2 + 0.5\epsilon
\end{equation*}

The results of case 1-3 are shown in following table and figures:
\begin{table}[ht]
  \centering
  \begin{tabular}{c c c c c c}
    \hline\hline
    ~ & p & S-SIR & KSIR & GS-KSIR-I & GS-KSIR-II\\
    \hline
    \multirow{2}{1.2cm}{Case 1}  & 10 & 0.3011(0.1199)  & 0.4613(0.1726)  & 0.7626(0.1244)  & 0.7817(0.1106)  \\
      & 20 & 0.2692(0.0793)  & 0.4592(0.1335)  & 0.7484(0.1615)  & 0.7305(0.1797)  \\
    \hline
    \multirow{2}{1.2cm}{Case 2}  & 10 & 0.0217(0.0145)  &  0.7508(0.0721) & 0.7546(0.0843)  & 0.7520(0.0832)  \\
      & 20 &  0.0236(0.0151)  & 0.7015(0.0550)  & 0.7038(0.0712)  & 0.6986(0.0854)  \\
    \hline
    \multirow{2}{1.2cm}{Case 3}  & 10 & 0.8172(0.1123)  & 0.6587(0.1360)  & 0.8124(0.1139)  & 0.8093(0.1149)  \\
      & 20 & 0.7583(0.1125)  & 0.6572(0.1328)  & 0.7927(0.1242)  &  0.7943(0.1262)  \\
    \hline
  \end{tabular}
  \caption{Mean and standard deviation of $\bar{r}^2$ in Case 1-3} \label{table01}
\end{table}

The result of three cases are in the tables \ref{table01} and figures 1-3. Through the simulations, we find that GS-KSIR-I and GS-KSIR-II are almost the same. This is because GS-KSIR-I and GS-KSIR-II are both extended from SIR and can be seen as a special case of the general semiparametric approach. Hence we consider GS-KSIR-I and GS-KSIR-II as GS-KSIR to compare K-SIR and S-SIR in the following context. Because GS-KSIR and K-SIR are designed for nonlinear dimension reduction structrue, it is reasonable that GS-KSIR and KSIR outperform S-SIR in the case of nonlinear dimension reduction structure. while in the case of linear dimension reduction structure, the performance of GS-KSIR and S-SIR is similar. As our method do not rely on linearity condition and constant variance condition, we can see that GS-KSIR outperforms KSIR in case 1. Then it is shown that GS-KSIR not only inherit the good properties of semiparametric linear SDR, but also expand the scope of the linear SDR.

\section{Empirical Application \label{real_sec}}
\begin{figure}[h!t]\label{figure04}
  \centering
  \includegraphics[width=0.8\textwidth]{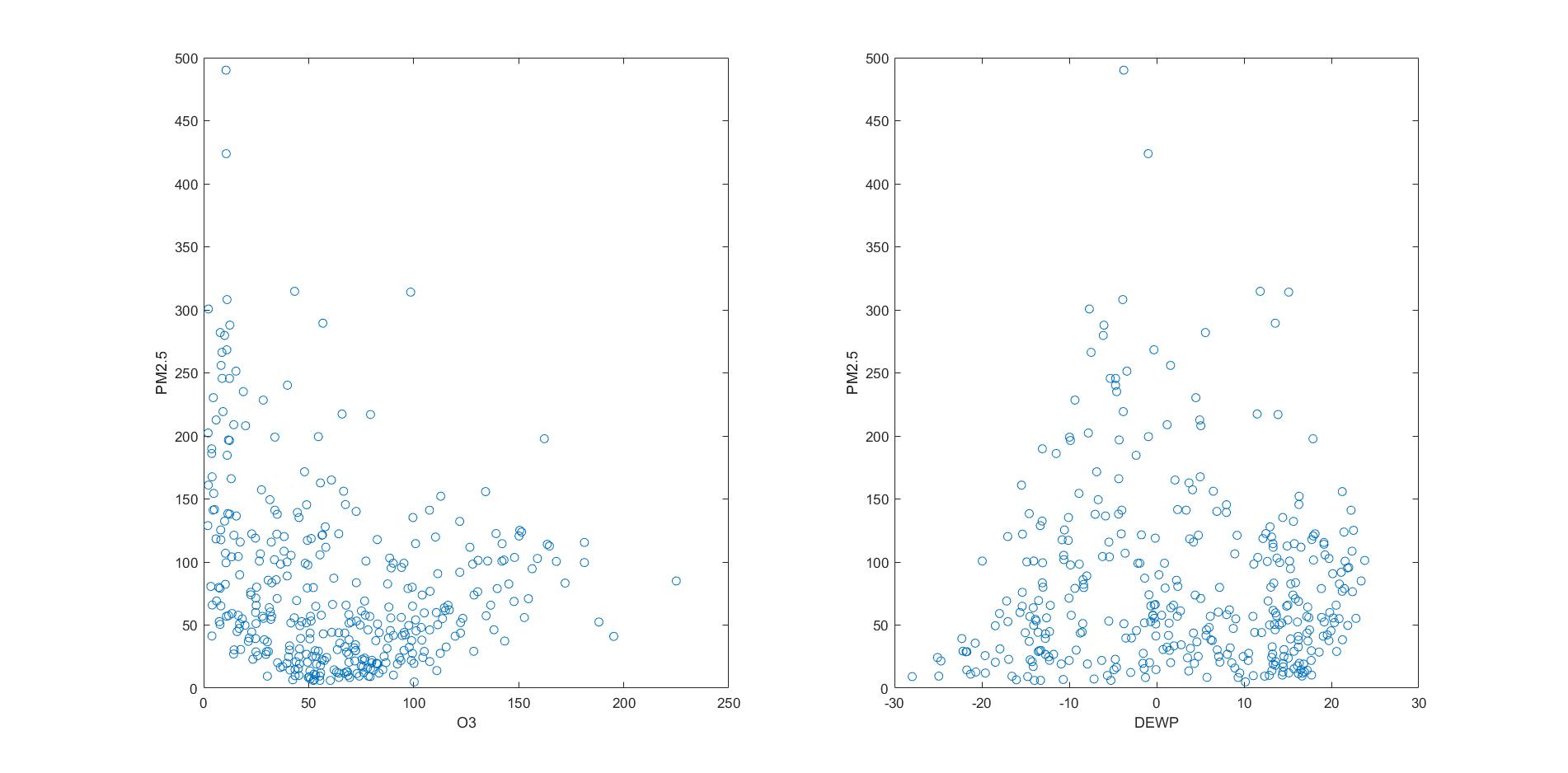}
  \caption{scatter plots of daily mean $PM_{2.5}$ against mean $O_{3}$ concentration($ug/m^3$) and mean dew point temperature(degree Celsius)}
\end{figure}

In this section, we use the data set which includes hourly air pollutants data from 12 nationally-controlled air-quality monitoring sites in Beijing for March 1st, 2013 to February 28th, 2017. This data set is available in the UCI machine learning repository. We consider the data for the year 2015 in the Aotizhongxin site and explore the relationship between the daily mean $PM_{2.5}$ concentration and some predictors. In this study, the number of observations $n = 365$ and there are ten explanatory variables that are possible associated with the daily mean $PM_{2.5}$ concentration: mean $PM_{10}$ concentration($ug/m^3$), mean $SO_{2}$ concentration($ug/m^3$), mean $NO_{2}$  concentration($ug/m^3$), mean $CO$  concentration($ug/m^3$), mean $O_{3}$   concentration($ug/m^3$), mean temperature(degree Celsius), mean pressure(hPa), mean dew point temperature(degree Celsius), mean wind speed($m/s$) and a binary variable indicating whether is rainy. The scatter plots of daily mean $PM_{2.5}$ against mean $O_{3}$ concentration($ug/m^3$) and mean dew point temperature(degree Celsius) in figure \ref{figure04} show that there exists nonlinearity in the data. Hence it is more appropriate to perform a nonlinear dimension reduction method than linear dimension reduction method on this data.

In this study we try to reduce the dimension of covariates to $d=2$. After reducing the dimension of covariates to 2, we choose gaussian process regression to fit the nonparametric function on $(\widehat{\bm{U}}(\bm{x}_{i}), \bm{Y}_i), i = 1, \ldots, n$ and perform a k-folds cross validation procedure to calculate the prediction mean absolute errors(PMAE). The prediction errors of Semi-SIR, KSIR, GS-KSIR-I are 12.9861, 5.8994 and 5.5266 respectively. From the above results, we can find out that GS-KSIR-I perform a litter better than KSIR, while both GS-KSIR-I and KSIR performance better than Semi-SIR. So, it is reasonable to consider that the dimension reduction structure of the data is nonlinear. GS-KSIR-I method has found a good nonlinear transform of the covariates X, which establish the relationship between the daily mean $PM_{2.5}$ concentration and some characteristics of air index.

\section{Conclusion \label{con_sec}}

In this paper, we have introduced a new nonlinear dimension reduction method which do not rely on the linearity condition and constant variance condition. We extended the semiparametric method to handle the situation where both the interested parameters and nuisance parameters are infinite dimensional. The generalized nuisance tangent space and its orthogonal complement are derived. The penalized estimation equation were constructed from the generalized nuisance tangent space
orthogonal complement. We proved the consistency of the estimator from the penalized estimation equation. Generalized semiparametric kernel sliced inverse regression proposed in this paper can be derived through two perspectives. The two methods perform almost same in the simulation studies. Specially, both of them perform better than KSIR when the covariate X violates the linearity condition and constant variance condition. The results of the simulation and real data studies present the effectiveness of our method. It has been shown that our method not only inherits the good properties of the method in \cite{MaZhu12}, which do not need the assumption of linearity and/or constant variance on the covariates, but also widens the application scope of the semiparametric dimension reduction method. With the framework of generalized semiparametric model, more methods like GS-DR and GS-PHD can be obtained through the similar derivation process. Moreover, \cite{MaZhu12} mentioned the sparsity assumption in handling the situation that p is very large in comparison with sample n. It is still a meaningful subject in the generalized semi-parametric frame work and also deserves future work.

\section{Reference}
\bibliographystyle{plain}
\bibliography{gsksirref}

\begin{thebibliography}{10}

\bibitem{akaho2006kernel}
Shotaro Akaho.
\newblock A kernel method for canonical correlation analysis.
\newblock {\em arXiv preprint cs/0609071}, 2006.

\bibitem{Altman1957}
M.~Altman.
\newblock A fixed point theorem in hilbert space.
\newblock {\em Bull. Polish Acad. Sci}, 5:19--22, 1957.

\bibitem{BachJordan2002}
Francis~R Bach and Michael~I Jordan.
\newblock Kernel independent component analysis.
\newblock {\em Journal of machine learning research}, 3(Jul):1--48, 2002.

\bibitem{Bickel1993}
Peter~J Bickel, Chris~AJ Klaassen, Peter~J Bickel, Y~Ritov, J~Klaassen, Jon~A
  Wellner, and YA'Acov Ritov.
\newblock {\em Efficient and adaptive estimation for semiparametric models},
  volume~4.
\newblock Johns Hopkins University Press Baltimore, 1993.

\bibitem{Cook1998}
R~Dennis Cook.
\newblock {\em Regression graphics: ideas for studying regressions through
  graphics}, volume 482.
\newblock John Wiley \& Sons, 2009.

\bibitem{CookandLi2002}
R~Dennis Cook, Bing Li, et~al.
\newblock Dimension reduction for conditional mean in regression.
\newblock {\em The Annals of Statistics}, 30(2):455--474, 2002.

\bibitem{Cook1991}
R~Dennis Cook and Sanford Weisberg.
\newblock Discussion of “sliced inverse regression for dimension
  reduction”.
\newblock {\em Journal of the American Statistical Association}, 86(414):335,
  1991.

\bibitem{LiandDong2010}
Yuexiao Dong and Bing Li.
\newblock Dimension reduction for non-elliptically distributed predictors:
  second-order methods.
\newblock {\em Biometrika}, 97(2):279--294, 2010.

\bibitem{fukumizu2007statistical}
Kenji Fukumizu, Francis~R Bach, and Arthur Gretton.
\newblock Statistical consistency of kernel canonical correlation analysis.
\newblock {\em Journal of Machine Learning Research}, 8(Feb):361--383, 2007.

\bibitem{Kenji2009}
Kenji Fukumizu, Francis R.Bach, and Michael I.Jordan.
\newblock Kernel dimension reduction in regression[j].
\newblock {\em The Annals of Statistics}, 37(4):1871--1905, 2009.

\bibitem{Lee2013}
Kuang-Yao Lee, Bing Li, Francesca Chiaromonte, et~al.
\newblock A general theory for nonlinear sufficient dimension reduction:
  Formulation and estimation.
\newblock {\em The Annals of Statistics}, 41(1):221--249, 2013.

\bibitem{LiandDong2009}
Bing Li, Yuexiao Dong, et~al.
\newblock Dimension reduction for nonelliptically distributed predictors.
\newblock {\em The Annals of Statistics}, 37(3):1272--1298, 2009.

\bibitem{LiandWang2007}
Bing Li and Shaoli Wang.
\newblock On directional regression for dimension reduction.
\newblock {\em Journal of the American Statistical Association},
  102(479):997--1008, 2007.

\bibitem{Li1991}
Ker-Chau Li.
\newblock Sliced inverse regression for dimension reduction.
\newblock {\em Journal of the American Statistical Association},
  86(414):316--327, 1991.

\bibitem{Li1992}
Ker-Chau Li.
\newblock On principal hessian directions for data visualization and dimension
  reduction: Another application of stein's lemma.
\newblock {\em Journal of the American Statistical Association},
  87(420):1025--1039, 1992.

\bibitem{Li2008}
Lexin Li and Xiangrong Yin.
\newblock Sliced inverse regression with regularizations.
\newblock {\em Biometrics}, 64(1):124--131, 2008.

\bibitem{maunderstanding}
Yanyuan Ma, Fei Jiang, and Masayuki Henmi.
\newblock Understanding and utilizing the linearity condition in dimension
  reduction.

\bibitem{ma2012efficiency}
Yanyuan Ma and LIPING Zhu.
\newblock Efficiency loss caused by linearity condition in dimension reduction.
\newblock {\em Biometrika}, 99(1):1--13, 2012.

\bibitem{MaZhu12}
Yanyuan Ma and Liping Zhu.
\newblock A semiparametric approach to dimension reduction.
\newblock {\em Journal of the American Statistical Association},
  107(497):168--179, 2012.

\bibitem{ma2013efficient}
Yanyuan Ma and Liping Zhu.
\newblock Efficient estimation in sufficient dimension reduction.
\newblock {\em Annals of statistics}, 41(1):250, 2013.

\bibitem{Tsiatis2006}
A.~A. Tsiatis.
\newblock {\em Semiparametric Theory and Missing Data}.
\newblock New York: Springer, 2007.

\bibitem{Wu08}
Han-Ming Wu.
\newblock Kernel sliced inverse regression with applications to classification.
\newblock {\em Journal of Computational and Graphical Statistics},
  17(3):590--610, 2008.

\bibitem{Wu2013}
Qiang Wu, Feng Liang, and Sayan Mukherjee.
\newblock Kernel sliced inverse regression: regularization and consistency.
\newblock In {\em Abstract and Applied Analysis}, volume 2013. Hindawi, 2013.

\bibitem{Xia2007}
Yingcun Xia.
\newblock A constructive approach to the estimation of dimension reduction
  directions.
\newblock {\em The Annals of Statistics}, 35(6):2654--2690, 2007.

\bibitem{Xia2002}
Yingcun Xia, Howell Tong, Wai~Keungxs Li, and Li-Xing Zhu.
\newblock An adaptive estimation of dimension reduction space.
\newblock {\em Journal of the Royal Statistical Society: Series B (Statistical
  Methodology)}, 64(3):363--410, 2002.

\bibitem{HuangLee2009}
Yi-Ren Yeh, Su-Yun Huang, and Yuh-Jye Lee.
\newblock Nonlinear dimension reduction with kernel sliced inverse regression.
\newblock {\em IEEE Transactions on Knowledge and Data Engineering},
  21(11):1590--1603, 2008.

\bibitem{Zhong2005}
Wenxuan Zhong, Peng Zeng, Ping Ma, Jun~S Liu, and Yu~Zhu.
\newblock Rsir: regularized sliced inverse regression for motif discovery.
\newblock {\em Bioinformatics}, 21(22):4169--4175, 2005.

\bibitem{zhu1996asymptotics}
Li-Xing Zhu, Kai-Tai Fang, et~al.
\newblock Asymptotics for kernel estimate of sliced inverse regression.
\newblock {\em The Annals of Statistics}, 24(3):1053--1068, 1996.

\end{thebibliography}
\clearpage
\appendix{}
\section{}

\textbf{A.1. Theory of Reproducing Kernel Hilbert Space}

Denote $\mathcal{H}_R$ as a Hilbert function space defined on $\mathcal{X}$, a subset of $\mathbb{R}^p$. Furthermore $\mathcal{H}_R$ is a RKHS if there is a two-dimensional function $R(s,t),\forall s,t\in\mathcal{X}$ satisfying:
\begin{itemize}
  \item[(1)] $R_s(t)=R(s,t)$ as a function of $t$ meets that for any $s\in \mathcal{X}$, $R_s(t)\in \mathcal{H}_R$
  \item[(2)] reproducing property holds which is that for every $s\in \mathcal{X}$ and $f\in \mathcal{H}_R$, $f(s)=\langle f,R(s,\cdot)\rangle$.
\end{itemize}
The two-dimensional function $R(s,t)$ is called the reproducing kernel of $\mathcal{H}_R$ and it could be uniquely determined by $\mathcal{H}_R$. If $\mathcal{H}_R$ is a separable Hilbert space with $R(s, t)$ as its reproducing kernel and $\{\varphi_j\}_{j=1}^{\infty}$ as its standard orthogonal basis, then the corresponding $R$ has its spectral decomposition:
{\begin{equation*}
  R(s,t)=\sum\limits_{j=1}^{\infty}\lambda_j\varphi_j(s)\varphi_j(t),\,\forall s,t\in \mathcal{X}
\end{equation*}}
Define the map $\phi:\mathcal{X}\mapsto L_2$ satisfies
{\begin{equation*}
  \phi(s)=(\sqrt{\lambda_1}\varphi_1(s),\sqrt{\lambda_2}\varphi_2(s)),\cdots)^T,\,\forall s\in \mathcal{X}
\end{equation*}}
Obviously, $\phi$ is a mapping between $\mathcal{X}$ and the space $L_2$. Define the inner product:
{\begin{equation*}
  \langle \phi(s),\phi(t)\rangle=\phi(s)^T\phi(t)=R(s,t),\,\forall s,t\in \mathcal{X}
\end{equation*}}
Let $\mathcal{H}$ be the completion of the space spanned by $\{\phi(s),\,\forall s\in \mathcal{X}\}$ and construct a map $\tau:\mathcal{H}_R\mapsto \mathcal{H}$ so that:
{\begin{equation*}
  \tau(R_s(\cdot))=\tau(R(s,\cdot))=\phi(s),\,\forall s\in\mathcal{X}
\end{equation*}
\begin{equation*}
  \left<\tau(R_s(\cdot)),\tau(R_t(\cdot))\right>_{\mathcal{H}_R}=\left<\phi(s),\phi(t)\right>_{\mathcal{H}}=R(s,t),\,\forall s,t\in \mathcal{X}
\end{equation*}}
Then it is not hard to see that $\tau$ is an isometric isomorphism mapping.

\textbf{A.2. Derivation of equation (\ref{(10)})}

For $\forall i = 1, \ldots, n$, we have
\begin{equation*}
  \hat{\xi}(y_i) = \frac{\sum_{j=1}^{n}\phi(x_j)K_{h_1}^{(1)}(y_j, y_i)}{\sum_{j=1}^{n}K_{h_1}^{(1)}(y_j, y_i)} =
  \mathbf{\Phi}\begin{pmatrix}
                 \frac{K_{h_1}^{(1)}(y_1, y_i)}{\sum_{j=1}^{n}K_{h_1}^{(1)}(y_j, y_i)} \\
                 \vdots \\
                 \frac{K_{h_1}^{(1)}(y_n, y_i)}{\sum_{j=1}^{n}K_{h_1}^{(1)}(y_j, y_i)}
               \end{pmatrix}
\end{equation*}

\begin{equation*}
  \hat{\zeta}(\beta^T\phi(x_i)) = \sum_{j=1}^{n}\hat{\xi}(y_i)\frac{K_{h_2}^{(2)}(\beta^T\phi(x_j), \beta^T\phi(x_i))}{\sum_{j=1}^{n}K_{h_2}^{(2)}(\beta^T\phi(x_j), \beta^T\phi(x_i))} = \mathbf{\Phi}\tilde{K}_{h_1}^{(1)}\begin{pmatrix}
                                      \frac{K_{h_2}^{(2)}(\beta^T\phi(x_j), \beta^T\phi(x_1))}{\sum_{j=1}^{n}K_{h_2}^{(2)}(\beta^T\phi(x_j), \beta^T\phi(x_i))} \\
                                      \vdots \\
                                      \frac{K_{h_2}^{(2)}(\beta^T\phi(x_n), \beta^T\phi(x_i))}{\sum_{j=1}^{n}K_{h_2}^{(2)}(\beta^T\phi(x_j), \beta^T\phi(x_i))}
                                    \end{pmatrix}
\end{equation*}

\begin{equation*}
  \hat{\theta}(\beta^T\phi(x_i)) = \sum_{j=1}^{n}\phi(x_j)\frac{K_{h_2}^{(2)}(\beta^T\phi(x_j), \beta^T\phi(x_i))}{\sum_{j=1}^{n}K_{h_2}^{(2)}(\beta^T\phi(x_j), \beta^T\phi(x_i))} = \mathbf{\Phi}\begin{pmatrix}
                                      \frac{K_{h_2}^{(2)}(\beta^T\phi(x_1), \beta^T\phi(x_i))}{\sum_{j=1}^{n}K_{h_2}^{(2)}(\beta^T\phi(x_j), \beta^T\phi(x_i))} \\
                                      \vdots \\
                                      \frac{K_{h_2}^{(2)}(\beta^T\phi(x_n), \beta^T\phi(x_i))}{\sum_{j=1}^{n}K_{h_2}^{(2)}(\beta^T\phi(x_j), \beta^T\phi(x_i))}
                                    \end{pmatrix}
\end{equation*}

Substitute the above formula into equation (\ref{(9)}), we can get
\begin{align*}
  \sum_{i=1}^{n}\left[ \hat{\xi}(y_i)(\phi(x_i) - \hat{\theta}(\beta^T\phi(x_i)))^{T}\right] & = (\hat{\xi}(y_1), \ldots, \hat{\xi}(y_n))\begin{pmatrix}
                                            (\phi(x_1) - \hat{\theta}(\beta^T\phi(x_1)))^T \\
                                            \vdots \\
                                            (\phi(x_n) - \hat{\theta}(\beta^T\phi(x_n)))^T
                                          \end{pmatrix}\\
  ~ & = \mathbf{\Phi}\hat{\mathbf{K}}_{h_1}^{(1)}[\mathbf{\Phi}^T - (\hat{\mathbf{K}}_{h_2}^{(2)})^T\mathbf{\Phi}^T] \\
  ~ & = \mathbf{\Phi}\hat{\mathbf{K}}_{h_1}^{(1)}(\mathbf{I}_n - (\hat{\mathbf{K}}_{h_2}^{(2)})^T)\mathbf{\Phi}^T
\end{align*}
and
\begin{equation*}
  \sum_{i=1}^{n}\left[ \hat{\zeta}(\beta^T\phi(x_i))(\phi(x_i) - \hat{\theta}(\beta^T\phi(x_i)))^T\right] = \mathbf{\Phi}\hat{\mathbf{K}}_{h_1}^{(1)}\hat{\mathbf{K}}_{h_2}^{(2)}(\mathbf{I}_n - (\hat{\mathbf{K}}_{h_2}^{(2)})^T)\mathbf{\Phi}^T
\end{equation*}
Then
\begin{equation*}
    \frac{1}{n}\sum_{i=1}^{n}\left[\left(\hat{\xi}(y_i) - \hat{\zeta}(\bm{\beta}^{T}\phi(x_i))\right) \times \left(\phi(x_i) - \hat{\theta}(\bm{\beta}^{T}\phi(x_i))\right)\right] = 0
\end{equation*}
can be converted to
\begin{equation*}
  \bm{\Phi} \tilde{\mathbf{K}}_{h_1}^{(1)}(\mathbf{I}_n - \tilde{\mathbf{K}}_{h_2}^{(2)})(\mathbf{I}_n - (\tilde{\mathbf{K}}_{h_2}^{(2)})^{T})\bm{\Phi}^{T} = 0
\end{equation*}

\textbf{A.3. Proof of Lemma \ref{lemma2.1}}

First, we show that the space $\Lambda_1 = \{\text{mean-square closure of all } \Lambda_{\gamma_1}\}$ consists of all q-dimensional mean-zero functions of X.

Here we denote $\Lambda_{1s} = \left\{f(X): E(f) = 0 \right\}$.
For $\forall g \in \Lambda_{\gamma_1}$, because $\int f_X(x, \gamma_1)dx = 1$ implies $$\frac{\partial }{\partial \gamma_1}\int f_X(x, \gamma_1)dx = 0$$ for any x and $\gamma_1$,
\begin{equation*}
  E(S_{\gamma_1}(X)) = \int \frac{\partial f_X(x, \gamma_1)/\partial \gamma_1}{f_X(x, \gamma_1)}f_X(x, \gamma_1)dx = 0
\end{equation*}
Hence, we have
\begin{equation*}
  E(g) = E(B^{q \times r_1}S_{\gamma_1}(X)) = 0
\end{equation*}
which means that
\begin{equation}\label{(A.17)}
  g \in \Lambda_1
\end{equation}
On the other hand, for $\forall g \in \Lambda_{1s}$, consider the parametric submodel with density $f_X(x,\gamma_1) = f_0(x)(1 + \gamma_1^Tg(x))$, $\gamma_1$ is taken sufficiently small so that $(1 + \gamma_1^Tg(x)) \ge 0$ for all $x$.
The function $f_X(x, \gamma_1)$ is a density function since $f_X(x,\gamma_1) \ge 0$ and
\begin{equation*}
\begin{split}
   \int f_X(x, \gamma_1)dx & = \int f_0(x)(1 + \gamma_1^Tg(x))dx \\
     & = \int f_0(x)dx + \int \gamma_2^Tg(x)f_0(x)dx \\
     & = 1 + 0 =1
\end{split}
\end{equation*}
For this parametric submodel, the score function is
\begin{equation*}
\begin{split}
   S_{\gamma_1}(x) & = \frac{\partial \log f_0(x)(1 + \gamma_1^Tg(x))}{\partial \gamma_1}\Bigg |_{\gamma_1 = 0} \\
     & = g(x)
\end{split}
\end{equation*}
So, $g(x)$ is an element of this particular parametric submodel nuisance tangent space. Because if g(x) is the limit of $g_i(x) (g_i(x) \in \Lambda_{1s})$,then
\begin{equation*}
  E(g) = E(g - g_i + g_i) = E(g_i) + E(g - g_i) = 0
\end{equation*}
Hence all elements of $\Lambda_{1s}$ are either elements of a parametric submodel nuisance tangent space or a limit of such elements, i.e.
\begin{equation}\label{(A.18)}
  \Lambda_{1s} \subset \Lambda_1
\end{equation}
From (\ref{(A.17)}) and (\ref{(A.18)}), we have
\begin{equation}\label{(A.19)}
  \Lambda_{1} = \left\{f(X): E(f) = 0 \right\}
\end{equation}

Next, we show that the space $\Lambda_2 = \{\text{mean-square closure of all } \Lambda_{\gamma_2}\}$ is the space of all q-dimensional random functions $f(Y,X)$ that satisfy $E(f|X) = E(f|U(X)) = 0$

Here we denote $\Lambda_{2s} = \left\{f: E(f|X)=E(f|U(X))= 0 \right\}$. For $\forall g \in \Lambda_{\gamma_2}$, because $$\int f_{Y|X}(y|X, \gamma_2)dy = \int f_{Y|U(X)}(y|U(X), \gamma_2)dy = 1$$ implies $$\frac{\partial }{\partial \gamma_2}f_{Y|X}(Y|X, \gamma_2)dy = 0$$ for all x and $\gamma_2$,
\begin{equation*}
\begin{split}
   E(S_{\gamma_2}(X,Y)|X) & = \int \frac{\partial f_{Y|X}(y|x, \gamma_{20})/\partial \gamma_2}{f_{Y|X}(y|x, \gamma_{20})}f_{Y|X}(y|x, \gamma_{20})dy \\
     & = \int \frac{\partial f_{Y|U(X)}(y|U(x), \gamma_{20})/\partial \gamma_2}{f_{Y|U(X)}(y|U(x), \gamma_{20})}f_{Y|U(X)}(y|U(x), \gamma_{20})dy \\
     & = E(S_{\gamma_2}(X,Y)|U(X)) = 0
\end{split}
\end{equation*}
Hence, we have
\begin{equation*}
\begin{split}
   E(g|X) & = E(B^{q \times \gamma_2}S_{\gamma_2}(X,Y)|X) \\
     & = E(B^{q \times \gamma_2}S_{\gamma_2}(X,Y)|U(X)) \\
     & = E(g|U(X)) = 0
\end{split}
\end{equation*}
which means that
\begin{equation}\label{(A.20)}
  g \in \Lambda_2
\end{equation}

On the other hand, for $\forall g \in \Lambda_{2s}$, consider the parametric submodel with density
\begin{equation*}
  f_{Y|X}(y|x, \gamma_1) = f_{0Y|X}(y|x)(1 + \gamma_2^Tg(y,x))
\end{equation*}
$\gamma_2$ is chosen sufficiently small so that
\begin{equation*}
  (1 + \gamma_2^Tg(x,y)) \ge 0 \text{ for all } y,x.
\end{equation*}
Similar with the derivation of $\Lambda_1$, the function $f_{Y|X}(y|x, \gamma_1)$ is a density function.So for this parametric submodel, the score vector is
\begin{equation*}
  S_{\gamma_2}(y,x) = \frac{\partial f_{Y|X}(y|x, \gamma_{20})}{\partial \gamma_2} = g(x,y)
\end{equation*}
Because $F(Y|X) = F(Y|U(X))$, we have $g(x,y)$ is an element of this particular parametric submodel nuisance tangent space.Similar with the derivation of $\Lambda_1$, we can obtain
\begin{equation}\label{(A.21)}
  \Lambda_2 \subset \Lambda_{2s}
\end{equation}
From (\ref{(A.20)}) and (\ref{(A.21)}), we have
\begin{equation}\label{(A.22)}
  \Lambda_2 = \left\{f: E(f|X)=E(f|U(X))= 0 \right\}
\end{equation}

\textbf{A.4. Proof of Theorem \ref{thm1}}

We denote the nuisance tangent space corresponding to $\eta_1$ and $\eta_2$, respectively, $\Lambda_1$ and $\Lambda_2$. We have
\begin{equation*}
  \Lambda_1 = \left\{f(X): E(f) = 0 \right\}
\end{equation*}
\begin{equation*}
  \Lambda_2 = \left\{ f(Y,X): E(f|X) = E(f|U(X)) = 0, \forall f\right\}
\end{equation*}
Obviously, $\Lambda_1 \perp \Lambda_2$, so $\Lambda = \Lambda_1 \oplus \Lambda_2$.

For $\forall f \in \Lambda_1, g \in \{f(Y,X): E(f|X) = 0\}$, we have
\begin{align*}
  E(f^{T}(X)g(Y,X)) & = E\{E(f^T(X)g(Y,X)|X\} \\
  ~ & = E\{f^T(X)E(g(Y,X)|X)\} \\
  ~ & = 0
\end{align*}
So,
\begin{equation}\label{(A.1)}
  \{f(Y,X): E(f|X) = 0\} \subset \Lambda_1^\perp
\end{equation}
On the other hand, for $\forall f(X) \in \Lambda_1$, $E(g(Y, X)|X)$ should be zero to make $E\{E(f^T(X)|X)E(g(Y,X)|X)\} = 0$, then
\begin{equation}\label{(A.2)}
  \{f(Y,X): E(f|X) = 0\} \supset \Lambda_1^\perp
\end{equation}
From (\ref{(A.1)}) and (\ref{(A.2)}), we have
\begin{equation*}
  \Lambda_1^\perp = \{f(Y,X): E(f|X) = 0\}
\end{equation*}

Next, we show that $\Lambda_2^\perp = \{f(Y,X): E(f|U(X), Y) \text{ is a function of } U(X) \text{ only}\}$.

For $\forall f \in \Lambda_2, g \in \mathcal{G} := \{f(Y,X): E(f|U(X), Y) \text{ is a function of } U(X) \text{ only}\}$, we have
\begin{align*}
  E\{f^T(Y,U(X))g(Y,X)\} & = E\{E[f^T(f,U(X))g(Y,X)|U(X),Y]\} \\
  ~ & = E\{f^T(Y,U(X))E(g(Y,X)|U(X,Y))\} \\
  ~ & = E\{f^T(Y,U(X))h(U(X),Y)\} \\
  ~ & = E\{E[f^T(Y,U(X))h(U(X), Y)]|U(X)\}\\
  ~ & = E\{h^T(U(X), Y)E[f^T(Y,U(X))|U(X)]\} \\
  ~ & = 0
\end{align*}
where $h(U(X,Y)) = E(g(Y,X)|U(X,Y))$.Hence
\begin{equation}\label{(A.3)}
  \mathcal{G} \subset \Lambda_2^\perp
\end{equation}
On the other side, for $\forall f(Y,X) \in \Lambda_2^\perp$, we set $g = E(f|U(X), Y) - E(f|U(X))$. Obviously, $g \in \Lambda_2$, so
\begin{align*}
  E(g^Tf) & = E\{g^TE[f|U(X), Y]\} \\
  ~ & = Eg^Tg + E\{g^TE(f|U(X))\} \\
  ~ & = Eg^Tg + E\{E[g^TE(f|U(X))|U(X)]\} \\
  ~ & = Eg^Tg \\
  ~ & = 0
\end{align*}
Then $g = E(f|U(X), Y) - E(f|U(X)) = 0$, which means $E(f|U(X), Y) = E(f|U(X))$ and
\begin{equation}\label{(A.4)}
  \Lambda_2^\perp \subset \mathcal{G}
\end{equation}
From (\ref{(A.3)}) and (\ref{(A.4)}), we have
\begin{equation*}
  \Lambda_2^\perp = \mathcal{G} = \{f(Y,X): E(f|U(X), Y) \text{ is a function of } U(X) \text{ only}\}
\end{equation*}

Finally, we show that
\begin{align*}
  \Lambda^\perp & = \Lambda_1^\perp \cap \Lambda_2^\perp\\
  ~ & = \{f(Y,X) - E(f|U(X),Y): E(f|X) = E(f|U(X)), \forall f\}
\end{align*}
We now denote
$$\mathcal{F} := \{f(Y,X) - E(f|U(X),Y): E(f|X) = E(f|U(X)), \forall f\}$$
Obviously, for $\forall g \in \mathcal{F}$ we have $E(g|U(X),Y) = 0$, then $\mathcal{F} \subset \Lambda_2^\perp$. Because
\begin{align*}
  E\{f(Y,X) - E(f|U(X),Y)|X\} & = E\{f(Y,X)|X\} - E\{E[f|U(X),Y]|X\} \\
  ~ & = E(f|X) - E(f|U(X)) \\
  ~ & = 0
\end{align*}
$\mathcal{F} \subset \Lambda_1^\perp$ as well. Hence we have
\begin{equation}\label{(A.5)}
  \mathcal{F} \subset \Lambda^\perp
\end{equation}
On the other hand, for $\forall f \in \Lambda^\perp$, because $f \in \Lambda_2^\perp$, there exists function a(x) such that
\begin{equation*}
  E(f|U(X),Y) = a(U(X))
\end{equation*}
then,
\begin{align*}
  a(U(X)) & = E(f|U(X) = U(x), Y) \\
  ~ & = \frac{\int_{U(X) = U(x)}f(Y,X)\eta_1(X)\eta_2(Y|U(X))d\mu(X)}{\int_{U(X) = U(x)}\eta_1(X)\eta_2(Y|U(X))d\mu(X)} \\
  ~ & = \frac{\int_{U(X) = U(x)}f(Y,X)\eta_1(X)d\mu(X)}{\int_{U(X) = U(x)}\eta_1(X)d\mu(X)}
\end{align*}
So,
\begin{equation}\label{(A.6)}
  \begin{split}
  a(U(X)) & = \int a(U(X))\eta_2(Y|U(X))d\mu(Y) \\
  ~ & = \int \frac{\int_{U(X)=U(x)}f(Y,X)\eta_1(X)d\mu(X)}{\int_{U(X)=U(x)}\eta_1(X)d\mu(X)}\eta_2(Y|U(X))d\mu(Y) \\
  ~ & = \frac{\int\int_{U(X)=U(x)}f(Y,X)\eta_1(X)\eta_2(Y|U(X))d\mu(X)d\mu(Y)}{\int_{U(X)=U(x)}\eta_1(X)d\mu(X)} \\
  ~ & = \frac{\int_{U(X)=U(x)}[\int f(Y,X)\eta_2(Y|U(X))d\mu(Y)]\eta_1(X)d\mu(X)}{\int_{U(X)=U(x)}\eta_1(X)d\mu(X)}
  \end{split}
\end{equation}

Because $f \in \Lambda_1^\perp$, then
\begin{equation}\label{(A.7)}
  \int f(Y,X)\eta_2(Y|U(X))d\mu(Y) = E(f(Y|X)|U(X)) = 0
\end{equation}
From (\ref{(A.6)}) and (\ref{(A.7)}), we have $a(U(X)) = 0$. Thus, the elements in $\Lambda^\perp$ has the form $f(Y,X) - E(f|U(X,Y)$, so
\begin{equation}\label{(A.8)}
  \Lambda^\perp \subset \mathcal{F}
\end{equation}
From (\ref{(A.5)}) and (\ref{(A.8)}), we have
\begin{equation*}
  \Lambda^\perp = \mathcal{F} = \{f(Y,X) - E(f|U(X),Y): E(f|X) = E(f|U(X)), \forall f\}
\end{equation*}

\textbf{A.5. Proof of Theorem (\ref{thm2}})

Because $\tilde{\mathbf{\Phi}}^T\tilde{\mathbf{\Phi}}= \tilde{R}$, equation (\ref{10_1}) can be converted to equation (\ref{(12)}).

On the other hand, the spectral decomposition of $\tilde{\mathbf{\Phi}}$ is as follows:
\begin{align*}
  \tilde{\mathbf{\Phi}} & = UDV^T \\
  ~ & = (u_1, \ldots, u_q)\begin{pmatrix}
                            \bar{D}_{q \times q} & 0_{q \times (p-q)} \\
                            0_{(n-q) \times q} & 0_{(n-q) \times (n-q)}
                          \end{pmatrix}\begin{pmatrix}
                                         v_1 \\
                                         \vdots \\
                                         v_n
                                       \end{pmatrix}\\
  ~ & = \bar{U}\bar{D}\bar{V}^T
\end{align*}
where $\bar{U} = (u_1, \ldots, u_q), \bar{V} = (v_1, \ldots, v_q)$. Then
\begin{equation*}
  \tilde{R} = \tilde{\mathbf{\Phi}}^T\tilde{\mathbf{\Phi}} = (\bar{V}\bar{D}\bar{U}^T)(\bar{U}\bar{D}\bar{V}^T) = \bar{V}\bar{D}^2\bar{V}^T
\end{equation*}
So, equation (\ref{(12)}) can be converted to
\begin{equation}\label{(A.9)}
  \bar{V}\bar{D}^2\bar{V}^T\tilde{\mathbf{K}}_{h_1}^{(1)}(\mathbf{I}_n - \tilde{\mathbf{K}}_{h_2}^{(2)})(\mathbf{I}_n - (\tilde{\mathbf{K}}_{h_2}^{(2)})^{T})\bar{V}\bar{D}^2\bar{V}^T = 0
\end{equation}
Because $(\bar{D}^{-2}\bar{V}^T)\bar{V}\bar{D}^2\bar{V}^T = \bar{V}^T$ and $(\bar{V}\bar{D}^2\bar{V}^T)\bar{V}\bar{D}^{-2} = \bar{V}$, equation (\ref{(A.9)}) can be converted to
\begin{equation}\label{(A.10)}
  \bar{V}^T\tilde{\mathbf{K}}_{h_1}^{(1)}(\mathbf{I}_n - \tilde{\mathbf{K}}_{h_2}^{(2)})(\mathbf{I}_n - (\tilde{\mathbf{K}}_{h_2}^{(2)})^{T})\bar{V} = 0
\end{equation}
Hence,
\begin{equation}\label{(A.11)}
  \bar{U}\bar{D}\bar{V}^T\tilde{\mathbf{K}}_{h_1}^{(1)}(\mathbf{I}_n - \tilde{\mathbf{K}}_{h_2}^{(2)})(\mathbf{I}_n - (\tilde{\mathbf{K}}_{h_2}^{(2)})^{T})\bar{V}\bar{D}\bar{U}^T = \tilde{\mathbf{\Phi}}\tilde{\mathbf{K}}_{h_1}^{(1)}(\mathbf{I}_n - \tilde{\mathbf{K}}_{h_2}^{(2)})(\mathbf{I}_n - (\tilde{\mathbf{K}}_{h_2}^{(2)})^{T})\tilde{\mathbf{\Phi}}^T
\end{equation}
then we can obtain the equation (\ref{(12)}).

From the above derivation, we have that equation (\ref{(10)}) is equivalent to equation (\ref{(12)}).

\textbf{A.6. Generalized Semiparametric Kernel Sliced Average Variance Estimation}

In this section, we derive the Generalized Semiparametric Kernel Sliced Average Variance Estimation from (\ref{(7)}). we set $g_1(Y, \bm{\beta}^{T}\phi(x)) = 1 - cov(\phi(x)|Y)$, $g_2(Y,\bm{\beta}^{T}\phi(x)) = g_1(Y,\bm{\beta}^{T}\phi(x))E(\phi(x)|Y)$ and $\alpha_1(x) = -\phi(x)(\phi(x) - E(\phi(x)|\bm{\beta}^{T}\phi(x))^{T}$, $\alpha_2(x) = \phi^{T}(x)$. Then we can construct the elements in $\Lambda^{\perp}$, which is
$$\sum_{i=1}^{2}\left\{g_i(Y,\bm{\beta}^{T}\phi(x)) - E(g_i|\bm{\beta}^{T}\phi(x))\right\}\left\{\alpha_i(x) - E(\alpha_i(x)|\bm{\beta}^{T}\phi(x))\right\}$$
Then we can obtain the estimation equation of SAVE:
\begin{equation}\label{GSKSAVE-01}
  E\left( [1-cov(\phi(x)|Y)][\{\phi(x) - E(\phi(x)|Y)\}\{\phi(x) - E(\phi(x)|\bm{\beta}^{T}\phi(x))\}^{T} - cov(\phi(x)|\bm{\beta}^{T}\phi(x))]\right) = \bm{0}
\end{equation}
To be convenient, we set
$\xi(Y) = E(\phi(x)|Y)$, $\theta(\beta^T\phi(x)) = E(\phi(x)|\bm{\beta}^{T}\phi(x) \rangle)$, $\varphi(Y) = cov(\phi(x)|Y)$, $\nu(\beta^T\phi(x)) = cov(\phi(x)|\beta^T\phi(x))$ and their corresponded kernel estimation to be $\widehat{\xi}(Y)$, $\widehat{\theta}(\beta^T\phi(x))$, $\widehat{\varphi}(Y)$, $\widehat{\nu}(Y)$.

Similar with the derivation of GSKSIR, we obtain the version of (\ref{GSKSAVE-01}):
\begin{equation}\label{GSKSAVE-02}
  \frac{1}{n}\sum_{i=1}^{n}[1-\widehat{\varphi}(y_i)][(\phi(x_i) - \widehat{\xi}(y_i))(\phi(x_i) - \widehat{\theta}(\bm{\beta}^{T}\phi(x_i))) - \widehat{\nu}(\bm{\beta}^{T}\phi(x_i))] = 0
\end{equation}
where
\begin{equation*}
  \widehat{\xi}(y_i) = \frac{\sum_{j=1}^{n}\phi(x_j)K_{h_1}^{(1)}(y_j, y_i)}{\sum_{j=1}^{n}K_{h_1}^{(1)}(y_j, y_i)}
\end{equation*}

\begin{equation*}
  \widehat{\theta}(\beta^T\phi(x_i)) = \sum_{j=1}^{n}\phi(x_j)\frac{K_{h_2}^{(2)}(\beta^T\phi(x_j), \beta^T\phi(x_i))}{\sum_{j=1}^{n}K_{h_2}^{(2)}(\beta^T\phi(x_j), \beta^T\phi(x_i))} = \sum_{j=1}^{n}\phi(x_j)\frac{K_{h_2}^{(2)}(\bm{C}^{T}R_{(j)}, \bm{C}^{T}R_{(i)})}{\sum_{j=1}^{n}K_{h_2}^{(2)}(\bm{C}^{T}R_{(j)}, \bm{C}^{T}R_{(i)})}
\end{equation*}

\begin{align*}
  \widehat{\varphi}(y_i) & = \widehat{E}(\phi(x)\phi^{T}(x)|y_i) - \widehat{E}(\phi(x)|y_i)\widehat{E}(\phi^{T}(x)|y_i)\\
  ~ & = \frac{\sum_{j=1}^{n}R_{jj}K_{h_3}^{(1)}(y_j, y_i)}{\sum_{j=1}^{n}K_{h_3}^{(1)}(y_j, y_i)} - \widehat{\xi}(y_i)\cdot\widehat{\xi}(y_i)
\end{align*}

\begin{align*}
  \widehat{\nu}(\bm{\beta}^{T}\phi(x_i)) & = \widehat{E}(\phi(x)\phi^{T}(x)|\bm{\beta}^{T}\phi(x_i) - \widehat{E}(\phi(x)|\bm{\beta}^{T}\phi(x_i)\widehat{E}(\phi^{T}(x)|\bm{\beta}^{T}\phi(x_i)\\
  ~ & = \sum_{j=1}^{n}R_{jj}\frac{K_{h_4}^{(2)}(\bm{C}^{T}R_{(j)}, \bm{C}^{T}R_{(i)})}{\sum_{j=1}^{n}K_{h_4}^{(2)}(\bm{C}^{T}R_{(j)}, \bm{C}^{T}R_{(i)})} - \widehat{\theta}(\beta^T\phi(x_i)) \cdot \widehat{\theta}(\beta^T\phi(x_i))
\end{align*}
we can convert equation (\ref{GSKSAVE-02}) into
\begin{equation}\label{GSKSAVE-03}
  \left[\bm{D}_1 - diag\{\bm{R}\}\bm{K}^{(1)}_{h_3}\right]^{T}\left[\bm{D}_2 - diag\{\bm{R}\}\bm{K}^{(2)}_{h_4}\right] = 0
\end{equation}
where notation $diag(\bm{A})$ represents the vector composed by the diagnose elements of matrix $\bm{A}$,
$$
\bm{D}_1 = diag\left\{I_n + (\bm{K}^{(1)}_{h_1})^{T}\bm{R}\bm{K}^{(1)}_{h_1}\right\}
$$
and
$$
\bm{D}_2 = diag\left\{\bm{R} - \bm{R}\bm{K}^{(1)}_{h_1} - \bm{R}\bm{K}^{(2)}_{h_2} + (\bm{K}^{(1)}_{h_1})^{T}\bm{R}\bm{K}^{(2)}_{h_1} + (\bm{K}^{(2)}_{h_2})^{T}\bm{R}\bm{K}^{(2)}_{h_2}\right\}
$$

\textbf{A.7. Asymptotic Property of Estimator}
\flushleft \textbf{Condition}
\begin{itemize}
  \item[C1. ] $K(\cdot)$ is Lipschitz continue, has compact support. It satisfies
      \begin{equation*}
        \int K(u)du = 1, \int u^{i}K(u)du = 0,1 \le i \le m-1, 0 \ne \int u^{m}K(u)du < \infty
      \end{equation*}
      The q-dimensional kernel function is a product of q univariate kernel function, that is,
      \begin{equation*}
        K_h(\mathbf{u}) = \frac{K(\mathbf{u}/h)}{h^{q}} = \prod_{j=1}^{q}K_h(u_j)
      \end{equation*}
      where $\mathbf{u} = (u_1, \ldots, u_q)^T$
  \item[C2. ] The $m$th derivation of $r_1(\bm{z})$, $r_2(\bm{z})$ and $f(\bm{z})$ are locally Lipschitz continues, where $r_1(\bm{z}) = E\left\{a(x)|\bm{z}\right\}f(\bm{z})$ and $r_2(\bm{z}) = E\left\{g(Y,\bm{z})|\bm{z}\right\}f(\bm{z})$, $\bm{z} = \bm{\beta}^{T}\phi(x)$.
  \item[C3. ] The density function of x and $\bm{z}$, that is, $f_x(x)$ and $f_{\bm{z}}(\bm{z})$ are bounded from below and above. Each entry in the matrices $E\left\{a(x)a^{T}(x)|\bm{\beta}^{T}\phi(x)\right\}$ and $E\left\{g(Y, \bm{\beta}^{T}\phi(x))g^{T}(Y, \bm{\beta}^{T}\phi(x))|\bm{\beta}^{T}\phi(x)\right\}$ is locally Lipschitz continuous and bounded from above as a function of $\bm{\beta}^{T}\phi(x)$
  \item[C4. ] The bandwidth $h = O(n^{-k})$ for $1/(4m) < k < 1/(2q)$ and $\lim_{n \rightarrow \infty} \sqrt{n}\lambda_{n} \rightarrow \infty$, $\rho_{n} = o(n^{-s}), 0<s<1/4$.
\end{itemize}

\begin{lemma}\label{lemmaA6.1}
  Assume Condition (C1)-(C4) holds. Then there exists $\bm{\beta}$ such that
  \begin{equation}\label{A6.1.1}
    \begin{split}
     \sup_{\|\widehat{\bm{\beta}} - \bm{\beta}\| \le \rho_{n}}  & \quad \left\|\widehat{E}\left\{a(x)|\widehat{\bm{\beta}}^{T}\phi(x)\right\} - \widehat{E}\left\{a(x)|\bm{\beta}^{T}\phi(x)\right\} - E\left\{a(x)|\widehat{\bm{\beta}}^{T}\phi(x)\right\} + E\left\{a(x)|\bm{\beta}^{T}\phi(x)\right\}\right\| \\
     = \quad & \quad O_p(\rho_{n}h^{m} + \rho_{n}^{2}h^{-(q+1)}\log n)
    \end{split}
  \end{equation}
   and
  \begin{equation}\label{A6.1.2}
    \begin{split}
       \sup_{\|\widehat{\bm{\beta}} - \bm{\beta}\| \le \rho_{n}}  & \quad \left\|\widehat{E}\left\{g(Y,\widehat{\bm{\beta}}^{T}\phi(x))|\widehat{\bm{\beta}}^{T}\phi(x)\right\} - \widehat{E}\left\{g(Y,\bm{\beta}^{T}\phi(x))|\bm{\beta}^{T}\phi(x)\right\} - E\left\{g(Y,\widehat{\bm{\beta}}^{T}\phi(x))|\widehat{\bm{\beta}}^{T}\phi(x)\right\}\right.\\
          & \left.\quad + \quad E\left\{g(Y,\bm{\beta}^{T}\phi(x))|\bm{\beta}^{T}\phi(x)\right\}\right\| \\
        = \quad & \quad O_p(\rho_{n}h^{m} + \rho_{n}^{2}h^{-(q+1)}\log n)
     \end{split}
  \end{equation}
\end{lemma}
\textbf{Proof for Lemma:}

Because the proof of (\ref{A6.1.1}) and (\ref{A6.1.2}) are similar, we only show the proof of (\ref{A6.1.1}). First, consider that the kernel estimation of $\widehat{E}\left\{a(x)|\widehat{\bm{\beta}}^{T}\phi(x)\right\}$ is
\begin{equation}\label{A6.1.3}
  \widehat{E}\left\{a(x)|\widehat{\bm{\beta}}^{T}\phi(x)\right\} = \frac{\frac{1}{n}\sum_{i=1}^{n}K_{h}\left(\widehat{\bm{\beta}}^{T}\phi(x_i)- \widehat{\bm{\beta}}^{T}\phi(x)\right)a(x_i)}{\frac{1}{n}\sum_{i=1}^{n}K_{h}\left(\widehat{\bm{\beta}}^{T}\phi(x_i)- \widehat{\bm{\beta}}^{T}\phi(x)\right)}
\end{equation}
Next we inspect the numerator and denominator of $\widehat{E}\left\{a(x)|\widehat{\bm{\beta}}^{T}\phi(x)\right\}$ respectively.
Here we focus on the boundness of $ES_i^2$
\begin{equation}\label{A6.1.4}
  \begin{split}
     ES_i^2 & = \quad E\left[\left\{K_{h}\left(\widehat{\bm{\beta}}^{T}\phi(x_i)- \widehat{\bm{\beta}}^{T}\phi(x)\right) - K_{h}\left(\bm{\beta}^{T}\phi(x_i)- \bm{\beta}^{T}\phi(x)\right)\right\}^2a(x_i)a^{T}(x_i)\right] \\
       & = \quad E\left[\left\{K_{h}\left(\widehat{\bm{\beta}}^{T}\phi(x_i)- \widehat{\bm{\beta}}^{T}\phi(x)\right) - K_{h}\left(\bm{\beta}^{T}\phi(x_i)- \bm{\beta}^{T}\phi(x)\right)\right\}^2E\left\{a(x_i)a^{T}(x_i)|\widehat{\bm{\beta}}^{T}\phi(x_i),\bm{\beta}^{T}\phi(x_i)\right\}\right]\\
       & = \quad E\left[\left\{K_{h}\left(\widehat{\bm{\beta}}^{T}\phi(x_i)- \widehat{\bm{\beta}}^{T}\phi(x)\right) - K_{h}\left(\bm{\beta}^{T}\phi(x_i)- \bm{\beta}^{T}\phi(x)\right)\right\}^2E\left\{a(x_i)a^{T}(x_i)|\bm{\beta}^{T}\phi(x_i)\right\}\right]  \\
       & + \quad E\left\{K_{h}\left(\widehat{\bm{\beta}}^{T}\phi(x_i)- \widehat{\bm{\beta}}^{T}\phi(x)\right) - K_{h}\left(\bm{\beta}^{T}\phi(x_i)- \bm{\beta}^{T}\phi(x)\right)\right\}^2\\
       & \times \quad  \left[E\left\{a(x_i)a^{T}(x_i)|\widehat{\bm{\beta}}^{T}\phi(x_i),\bm{\beta}^{T}\phi(x_i)\right\} - E\left\{a(x_i)a^{T}(x_i)|\bm{\beta}^{T}\phi(x_i)\right\}\right]
  \end{split}
\end{equation}
where
\begin{equation*}
  S_i = \left\{K_{h}\left(\widehat{\bm{\beta}}^{T}\phi(x_i)- \widehat{\bm{\beta}}^{T}\phi(x)\right) - K_{h}\left(\bm{\beta}^{T}\phi(x_i)- \bm{\beta}^{T}\phi(x)\right)\right\}a(x_i)
\end{equation*}
Under the Condition (C3) that $E\left\{a(x_i)a^{T}(x_i)|\bm{\beta}^{T}\phi(x_i)\right\} \le M_1$,  $f_x(x_i) \le M_2$ and Condition (C1) that $K(\cdot)$ is Lipschitz continuous and has compact support, the first term of (\ref{A6.1.4}) has that
\begin{equation}\label{A6.1.5}
  \begin{split}
        & \quad E\left[\left\{K_{h}\left(\widehat{\bm{\beta}}^{T}\phi(x_i)- \widehat{\bm{\beta}}^{T}\phi(x)\right) - K_{h}\left(\bm{\beta}^{T}\phi(x_i)- \bm{\beta}^{T}\phi(x)\right)\right\}^2E\left\{a(x_i)a^{T}(x_i)|\bm{\beta}^{T}\phi(x_i)\right\}\right]\\
      \le & \quad M_1M_2\int \left\{K_{h}\left(\widehat{\bm{\beta}}^{T}\phi(x_i)- \widehat{\bm{\beta}}^{T}\phi(x)\right) - K_{h}\left(\bm{\beta}^{T}\phi(x_i)- \bm{\beta}^{T}\phi(x)\right)\right\}^2 dx_i \\
      = & \quad M_1M_2h^{-2q}\int \left\{K\left(\frac{\widehat{\bm{\beta}}^{T}\phi(x_i)- \widehat{\bm{\beta}}^{T}\phi(x)}{h}\right) - K\left(\frac{\bm{\beta}^{T}\phi(x_i)- \bm{\beta}^{T}\phi(x)}{h}\right)\right\}^2 dx_i \\
      \le & \quad M_1M_2M_3h^{-2(q+1)}\int\left\|(\widehat{\bm{\beta}} - \bm{\beta})^{T}(\phi(x_i)-\phi(x))\right\|^2dx_i \\
      \le & \quad M_1M_2M_3h^{-2(q+1)}\|\widehat{\bm{\beta}} - \bm{\beta}\|^2\int\|\phi(x_i)-\phi(x)\|^2dx_i\\
      = & \quad O\left(\rho_{n}^{2}h^{-2(q+1)}\right)
  \end{split}
\end{equation}

Under the Condition (C3) that $E\left\{a(x_i)a^{T}(x_i)|\widehat{\bm{\beta}}^{T}\phi(x_i)\right\}$ is locally Lipschitz-continuous, we have
\begin{equation*}
  \left\|E\left\{a(x_i)a^{T}(x_i)|\widehat{\bm{\beta}}^{T}\phi(x_i),\bm{\beta}^{T}\phi(x_i)\right\} - E\left\{a(x_i)a^{T}(x_i)|\bm{\beta}^{T}\phi(x_i)\right\}\right\| \le M_4\left\|(\widehat{\bm{\beta}} - \bm{\beta})^{T}\phi(x_i)\right\|
\end{equation*}
Then with the condition that $f_x(x_i) \le M_2$, the second equality of (\ref{A6.1.4}) has that
\begin{equation}\label{A6.1.6}
  \begin{split}
      & \quad E\left\{K_{h}\left(\widehat{\bm{\beta}}^{T}\phi(x_i)- \widehat{\bm{\beta}}^{T}\phi(x)\right) - K_{h}\left(\bm{\beta}^{T}\phi(x_i)- \bm{\beta}^{T}\phi(x)\right)\right\}^2\cdot \left[E\left\{a(x_i)a^{T}(x_i)|\widehat{\bm{\beta}}^{T}\phi(x_i),\bm{\beta}^{T}\phi(x_i)\right\} - E\left\{a(x_i)a^{T}(x_i)|\bm{\beta}^{T}\phi(x_i)\right\}\right]\\
      \le & \quad M_2M_4\int \left\{K_{h}\left(\widehat{\bm{\beta}}^{T}\phi(x_i)- \widehat{\bm{\beta}}^{T}\phi(x)\right) - K_{h}\left(\bm{\beta}^{T}\phi(x_i)- \bm{\beta}^{T}\phi(x)\right)\right\}^2\left\|(\widehat{\bm{\beta}} - \bm{\beta})^{T}\phi(x_i)\right\| dx_i \\
      = & \quad M_2M_4h^{-2(q+1)}\int \left\{K\left(\widehat{\bm{\beta}}^{T}\phi(x_i)- \widehat{\bm{\beta}}^{T}\phi(x)\right) - K\left(\bm{\beta}^{T}\phi(x_i)- \bm{\beta}^{T}\phi(x)\right)\right\}^2\left\|(\widehat{\bm{\beta}} - \bm{\beta})^{T}\phi(x_i)\right\| dx_i \\
      \le & \quad M_2M_4h^{-2(q+1)} \int \left\|(\widehat{\bm{\beta}} - \bm{\beta})^{T}(\phi(x_i) - \phi(x))\right\|^2\left\|(\widehat{\bm{\beta}} - \bm{\beta})^{T}\phi(x_i)\right\|dx_i \\
      = & \quad o\left(\rho_{n}^{2}h^{-2(q+1)}\right)
  \end{split}
\end{equation}
Hence with (\ref{A6.1.5}) and (\ref{A6.1.6}), we can obtain that (\ref{A6.1.4}) is bounded by $O\left(\rho_{n}^{2}h^{-2(q+1)}\right)$.

Then by using Chebyshev's inequality, we have
\begin{equation*}
  \left|\frac{1}{n}\sum_{i=1}^{n}S_i - ES_i\right| = O_p\left(\rho_{n}^{2}h^{-(q+1)}\right)
\end{equation*}
Moreover, we can use Theorem 37 in Pollard (1984, page 34) to prove that
\begin{equation}\label{A6.1.7}
   \sup_{\|\widehat{\bm{\beta}} - \bm{\beta}\| \le \rho_{n}}\left|\frac{1}{n}\sum_{i=1}^{n}S_i - ES_i\right| = O_p\left(\rho_{n}^{2}h^{-(q+1)}\log n\right)
\end{equation}

Next we prove that
\begin{equation}\label{A6.1.8}
  \sup_{\|\widehat{\bm{\beta}} - \bm{\beta}\| \le \rho_{n}}\left|E\left[\left\{K_{h}\left(\widehat{\bm{\beta}}^{T}\phi(x_i)- \widehat{\bm{\beta}}^{T}\phi(x)\right) - K_{h}\left(\bm{\beta}^{T}\phi(x_i)- \bm{\beta}^{T}\phi(x)\right)\right\}a(x_i)\right] - r_1(\widehat{\bm{\beta}}^{T}\phi(x_i)) + r_1(\bm{\beta}^{T}\phi(x_i))\right| = O\left(\rho_{n}h^{m}\right)
\end{equation}
where $r_1(\widehat{\bm{\beta}}^{T}\phi(x_i)) = E\left\{a(x_i)|\widehat{\bm{\beta}}^{T}\phi(x_i)\right\}f(\widehat{\bm{\beta}}^{T}\phi(x_i))$. By Taylor expansion, we have
\begin{align*}
    & \quad E\left\{K_{h}\left(\widehat{\bm{\beta}}^{T}\phi(x_i)- \widehat{\bm{\beta}}^{T}\phi(x)\right)a(x_i)\right\} - r_1(\widehat{\bm{\beta}}^{T}\phi(x_i))\\
  = & \quad E\left[K_{h}\left(\widehat{\bm{\beta}}^{T}\phi(x_i)- \widehat{\bm{\beta}}^{T}\phi(x)\right)E\left\{a(x_i)|\widehat{\bm{\beta}}^{T}\phi(x_i)\right\}\right] - r_1(\widehat{\bm{\beta}}^{T}\phi(x_i))\\
  = & \quad \int K_{h}\left(\widehat{\bm{\beta}}^{T}\phi(x_i)- \widehat{\bm{\beta}}^{T}\phi(x)\right)r_1(\widehat{\bm{\beta}}^{T}\phi(x_i))d(\widehat{\bm{\beta}}^{T}\phi(x_i)) - r_1(\widehat{\bm{\beta}}^{T}\phi(x_i)) \\
  = & \quad \int K(z_i)r_1(\widehat{\bm{\beta}}^{T}\phi(x)+hz_i)dz_i - r_1(\widehat{\bm{\beta}}^{T}\phi(x)) \\
  = & \quad \int K(z_i)(hz_i)^{m}\left\{r_1^{(m)}(\widehat{\bm{\beta}}^{T}\phi(x)+hz_i^{\star})\right\}/m!dz_i
\end{align*}
where $z_i^{\star}$ is between $\widehat{\bm{\beta}}^{T}\phi(x)$ and $\widehat{\bm{\beta}}^{T}\phi(x) + \widehat{\bm{\beta}}^{T}\phi(x_i)$. Let $z_i^{\star\star}$ is between $\bm{\beta}^{T}\phi(x)$ and $\bm{\beta}^{T}\phi(x) + \bm{\beta}^{T}\phi(x_i)$. With the local Lipschitz continuity of $r_1^{(m)}$ in Condition (C2), we have
\begin{align*}
    & \quad \sup_{\|\widehat{\bm{\beta}} - \bm{\beta}\| \le \rho_{n}}\left|E\left[\left\{K_{h}\left(\widehat{\bm{\beta}}^{T}\phi(x_i)- \widehat{\bm{\beta}}^{T}\phi(x)\right) - K_{h}\left(\bm{\beta}^{T}\phi(x_i)- \bm{\beta}^{T}\phi(x)\right)\right\}a(x_i)\right] - r_1(\widehat{\bm{\beta}}^{T}\phi(x_i)) + r_1(\bm{\beta}^{T}\phi(x_i))\right| \\
  =  & \quad \sup_{\|\widehat{\bm{\beta}} - \bm{\beta}\| \le \rho_{n}}\left|\int K(z_i)(hz_i)^{m}\frac{r_1^{(m)}(\widehat{\bm{\beta}}^{T}\phi(x) + hz_i^{\star}) - r_1^{(m)}(\bm{\beta}^{T}\phi(x) + hz_i^{\star\star})}{m!}dz_i\right| \\
  = & \quad \sup_{\|\widehat{\bm{\beta}} - \bm{\beta}\| \le \rho_{n}}\left|\int K(z_i)(hz_i)^{m}\frac{r_1^{(m)}(\widehat{\bm{\beta}}^{T}\phi(x))(1 + O(hz_i^{\star})) - r_1^{(m)}(\bm{\beta}^{T}\phi(x))(1 + O(hz_i^{\star\star}))}{m!}dz_i\right| \\
  = & \quad O\left(\rho_{n}h^{m}\right)
\end{align*}

Hence with (\ref{A6.1.7}) and (\ref{A6.1.8}), we have
\begin{equation}\label{A6.1.9}
  \begin{split}
    & \quad \sup_{\|\widehat{\bm{\beta}} - \bm{\beta}\| \le \rho_{n}}\left|\frac{1}{n}\sum_{i=1}^{n}\left\{K_{h}\left(\widehat{\bm{\beta}}^{T}\phi(x_i)- \widehat{\bm{\beta}}^{T}\phi(x)\right) - K_{h}\left(\bm{\beta}^{T}\phi(x_i)- \bm{\beta}^{T}\phi(x)\right)\right\}a(x_i) - r_1(\widehat{\bm{\beta}}^{T}\phi(x_i)) + r_1(\bm{\beta}^{T}\phi(x_i))\right| \\
   = & \quad O_p\left(\rho_{n}h^{m} + \rho_{n}^{2}h^{-(q+1)}\log n\right)
  \end{split}
\end{equation}
Similar with the proof of (\ref{A6.1.9}), we let $a(x_i) = 1$ and can obtain that
\begin{equation}\label{A6.1.10}
  \begin{split}
    & \quad \sup_{\|\widehat{\bm{\beta}} - \bm{\beta}\| \le \rho_{n}}\left|\frac{1}{n}\sum_{i=1}^{n}\left\{K_{h}\left(\widehat{\bm{\beta}}^{T}\phi(x_i)- \widehat{\bm{\beta}}^{T}\phi(x)\right) - K_{h}\left(\bm{\beta}^{T}\phi(x_i)- \bm{\beta}^{T}\phi(x)\right)\right\} - f(\widehat{\bm{\beta}}^{T}\phi(x_i)) + f(\bm{\beta}^{T}\phi(x_i))\right| \\
   = & \quad O_p\left(\rho_{n}h^{m} + \rho_{n}^{2}h^{-(q+1)}\log n\right)
  \end{split}
\end{equation}
Then we can obtain (\ref{A6.1.1}) by (\ref{A6.1.9}) and (\ref{A6.1.10}).

\begin{lemma}\label{lemmaA6.2}
  Assume Conditions (C1)-(C3) hold. Then
  \begin{align*}
    & \frac{1}{n}\sum_{i=1}^{n}\left(g(Y_i, \bm{\beta}^{T}\phi(x_i)) - E\{g(Y_i, \bm{\beta}^{T}\phi(x_i))|\bm{\beta}^{T}\phi(x_i)\}\right)\left(\widehat{E}\{a(x_i)|\bm{\beta}^{T}\phi(x_i)\} - E\{a(x_i)|\bm{\beta}^{T}\phi(x_i)\}\right) \\
    = & O_p\left\{1/(nh^{q/2}) + h^m/n^{1/2} + h^{2m} + \log^2n/(nh^q)\right\}
  \end{align*}
  and
  \begin{align*}
    ~ & \frac{1}{n}\sum_{i=1}^{n}\left(\widehat{E}\left\{g(Y_i, \bm{\beta}^{T}\phi(x_i))|\bm{\beta}^{T}\phi(x_i)\right\} - E\left\{g(Y_i, \bm{\beta}^{T}\phi(x_i))|\bm{\beta}^{T}\phi(x_i)\right\}\right)\left(a(x_i) - E\{a(x_i)|\bm{\beta}^{T}\phi(x_i)\}\right) \\
    = & O_p\left\{1/(nh^{q/2}) + h^m/n^{1/2} + h^{2m} + \log^2n/(nh^q)\right\}
  \end{align*}
\end{lemma}

\textbf{Proof for Lemma:}

Because the proof of these two equalities are similar, we only show the details of first one.

First, we set
\begin{align*}
  \epsilon^{(2)}_{i} & = g(Y_i, \bm{\beta}^{T}\phi(x_i)) - E\{g(Y_i, \bm{\beta}^{T}\phi(x_i))|\bm{\beta}^{T}\phi(x_i)\} \\
  \hat{r}_2(\bm{\beta}^{T}\phi(x_i)) & = \frac{1}{n-1}\sum_{j \ne i}K_h(\bm{\beta}^{T}\phi(x_j),\bm{\beta}^{T}\phi(x_i))a(x_j) \\
  \hat{f}(\bm{\beta}^{T}\phi(x_i)) & = \frac{1}{n-1}\sum_{j \ne i}K_h(\bm{\beta}^{T}\phi(x_j),\bm{\beta}^{T}\phi(x_i))
\end{align*}
Then rewrite the first equation, we have
\begin{align*}
    ~ & \quad \frac{1}{n}\sum_{i=1}^{n}\epsilon^{(2)}_{i}\left(\hat{E}\{a(x_i)|\bm{\beta}^{T}\phi(x_i)\} - E\{a(x_i)|\bm{\beta}^{T}\phi(x_i)\}\right) \\
  = & \quad \frac{1}{n}\sum_{i=1}^{n}\epsilon^{(2)}_{i}\left[\frac{\hat{r}_{2}(\bm{\beta}^{T}\phi(x_i))}{\hat{f(\bm{\beta}^{T}\phi(x_i))}} - \frac{r_2(\bm{\beta}^{T}\phi(x_i))}{f(\bm{\beta}^{T}\phi(x_i))}\right] \\
  = & \quad \frac{1}{n}\sum_{i=1}^{n}\epsilon^{(2)}_{i}\left[\frac{\hat{r}_{2}(\bm{\beta}^{T}\phi(x_i)) - r_2(\bm{\beta}^{T}\phi(x_i))}{f(\bm{\beta}^{T}\phi(x_i))}\right] - \frac{1}{n}\sum_{i=1}^{n}\epsilon^{(2)}_{i}\left[\frac{r_2(\bm{\beta}^{T}\phi(x_i))\left\{\hat{f}(\bm{\beta}^{T}\phi(x_i)) - f(\bm{\beta}^{T}\phi(x_i))\right\}}{f^2(\bm{\beta}^{T}\phi(x_i))}\right]\\
  - & \quad \frac{1}{n}\sum_{i=1}^{n}\epsilon^{(2)}_{i}\left[\frac{\left\{\hat{r}_2(\bm{\beta}^{T}\phi(x_i))- r_2(\bm{\beta}^{T}\phi(x_i))\right\}\left\{\hat{f}(\bm{\beta}^{T}\phi(x_i)) - f(\bm{\beta}^{T}\phi(x_i))\right\}}{f(\bm{\beta}^{T}\phi(x_i))\hat{f}(\bm{\beta}^{T}\phi(x_i))}\right] \\
  + & \quad \frac{1}{n}\sum_{i=1}^{n}\epsilon^{(2)}_{i}\left[\frac{r_2(\bm{\beta}^{T}\phi(x_i))\left\{\hat{f}(\bm{\beta}^{T}\phi(x_i)) - f(\bm{\beta}^{T}\phi(x_i))\right\}^2}{f^2(\bm{\beta}^{T}\phi(x_i))\hat{f}(\bm{\beta}^{T}\phi(x_i))}\right]
\end{align*}

By the uniform convergence of nonparametric regression, the third and fourth quantities are order $O_p(h^{2m} + \log^2n/(nh^q))$. Then we focus on the first and second quantities, these two quantities have similar structure, so we only study the first one.

We rewrite $n^{-1}\sum_{i=1}^{n}\hat{r}_2(\bm{\beta}^{T}\phi(x_i))\epsilon^{(2)}_{i}$ as a second order U-statistic, that is,
\begin{equation*}
  \frac{1}{n}\sum_{i=1}^{n}\hat{r}_2(\bm{\beta}^{T}\phi(x_i))\epsilon^{(2)}_{i} = \frac{1}{n(n-1)}\sum_{i \ne j}^{n}K_h(\bm{\beta}^{T}\phi(x_i), \bm{\beta}^{T}\phi(x_j))(\epsilon^{(2)}_{i}a(x_j) + \epsilon^{(2)}_{j}a(x_i))
\end{equation*}
With the Lemma 5.2.1.A of Serfling(1980, page 183), we have
\begin{equation*}
  \frac{1}{n}\sum_{i=1}^{n}\hat{r}_2(\bm{\beta}^{T}\phi(x_i))\epsilon^{(2)}_{i} - \frac{1}{n}\sum_{i=1}^{n}E\left\{K_h(\bm{\beta}^{T}\phi(x_i), \bm{\beta}^{T}\phi(x_j))r_2(\bm{\beta}^{T}\phi(x_j))|\bm{\beta}^{T}\phi(x_i)\right\}\epsilon^{(2)}_{i} = O_p\left(1/\left(nh^{q/2}\right)\right)
\end{equation*}
By using the condition that the $m$th derivative of $r_2(\bm{\beta}^{T}\phi(x))f(\bm{\beta}^{T}\phi(x))$ is locally Lipschitz-continuous and referring to the derivation of (\ref{A6.1.8}), we can obtain that
\begin{equation*}
  \sup_{x_i}\left|E\left\{K_h(\bm{\beta}^{T}\phi(x_i), \bm{\beta}^{T}\phi(x_j))r_2(\bm{\beta}^{T}\phi(x_j))|\bm{\beta}^{T}\phi(x_i)\right\} - r_2(\bm{\beta}^{T}\phi(x_i))f(\bm{\beta}^{T}\phi(x_i))\right| = O_p(h^m)
\end{equation*}
Further, we have
\begin{equation*}
  \frac{1}{n}\sum_{i=1}^{n}\left[E\left\{K_h(\bm{\beta}^{T}\phi(x_i), \bm{\beta}^{T}\phi(x_j))r_2(\bm{\beta}^{T}\phi(x_j))|\bm{\beta}^{T}\phi(x_i)\right\} - r_2(\bm{\beta}^{T}\phi(x_i))f(\bm{\beta}^{T}\phi(x_i))\right]\epsilon^{(2)}_{i} = O_p(h^m/n^{1/2})
\end{equation*}

Hence, we can obtain
\begin{equation*}
  \frac{1}{n}\sum_{i=1}^{n}\left[\hat{r}_2(\bm{\beta}^{T}\phi(x_i)) - r_2(\bm{\beta}^{T}\phi(x_i))f(\bm{\beta}^{T}\phi(x_i))\right]\epsilon^{(2)}_{i} = O_p\left(1/(nh^{q/2}) + h^m/n^{1/2}\right)
\end{equation*}

Finally, with the above derivation, we have
  \begin{align*}
    & \frac{1}{n}\sum_{i=1}^{n}\left(g(Y_i, \bm{\beta}^{T}\phi(x_i)) - E\{g(Y_i, \bm{\beta}^{T}\phi(x_i))|\bm{\beta}^{T}\phi(x_i)\}\right)\left(\hat{E}\{a(x_i)|\bm{\beta}^{T}\phi(x_i)\} - E\{a(x_i)|\bm{\beta}^{T}\phi(x_i)\}\right) \\
    = & O_p\left\{1/(nh^{q/2}) + h^m/n^{1/2} + h^{2m} + \log^2n/(nh^q)\right\}
  \end{align*}

\begin{lemma}\label{lemmaA6.4}
\textbf{(Leray-Schauder Theorem)}
Let $\Omega$ be an open, bounded set in a real-value Hilbert space $\mathcal{H}$, and assume that $F:\bar{\Omega} \rightarrow \mathcal{H}$ is continuous and satisfies $\inf_{x \in \partial\Omega}\langle x-x_0, Fx\rangle_{\mathcal{H}} \ge 0$, where $x_0 \in \Omega$. Then $Fx = 0$ has a solution in $\bar{\Omega}$.
\end{lemma}

\textbf{Proof for Lemma:}

Set $\Omega_0 = \{x|x+x_0 \in \Omega\}$ and define $G: \overline{\Omega}_{0} \rightarrow \mathcal{H}$ by $Gx = x - F(x+x_0)$. Obviously G is continuous. Now let $x \in \Omega_0$, then $x + x_0 \in \Omega$, therefore,
\begin{equation*}
  \langle x, x-Gx\rangle_{\mathcal{H}} = \langle x,F(x+x_0)\rangle_{\mathcal{H}} \ge 0
\end{equation*}
Then we obtain
$\langle x, Gx\rangle \le \|x\|^2$ for $\forall x \in \partial\Omega$. Then by the Altman theorem in \cite{Altman1957}, G has a fixed point $x^{\star} \in \bar{\Omega}_0$ so that $F(x^{\star} + x_0) = 0$.

Next, we consider the Hilbert Space $\mathcal{H}^{q}$ with the norm $\|x\|^2_{\mathcal{H}^{q}} = \sum_{i=1}^{q}\|x_i\|^2_{\mathcal{H}}$. Similarly, we can also obtain that $Fx = 0$ has a solution in $\bar{\Omega}$, where $\bar{\Omega}$ is an open, bounded set in $\mathcal{H}^{q}$.
~\\
\textbf{Taylor Expansion in Banach Space}\\
Suppose $E_1, E_2$ is Banach space, $D$ is the open subset in $E_1$, $L = \{x|x = x_0 + th, 0 \le t \le 1\} \subset E_1$ and $A : D \rightarrow E_2$. If $A^{(n)}(x)$ exists and is continue in $L$, then we have
\begin{equation*}
  A(x_0 + h) = Ax_0 + A^{\prime}(x_0)h + \ldots + \frac{1}{(n-1)!}A^{(n-1)}(x_0)h^{n-1} + \int_{0}^{1}(1-t)^{n-1}A^{(n)}(x_0+th)h^{n}dt
\end{equation*}
Specially, when $n = 1$, we have
\begin{equation*}
  \|A(x_0 + h) - Ax_0 - A^{\prime}(x_0)h\| \le \int_{0}^{1}\|[A^{\prime}(x_0+th) - A^{\prime}(x_0)]h\|dt
\end{equation*}
If $A^{\prime}(x)$ satisfies
\begin{equation*}
  \|A^{\prime}(x) - A^{\prime}(y)\| \le M_1\|x-y\|
\end{equation*}
then we have
\begin{equation*}
  \|A(x_0 + h) - Ax_0 - A^{\prime}(x_0)h\| \le \int_{0}^{1}tM_1\|h\|^2dt = \frac{1}{2}M_1\|h\|^2
\end{equation*}
\begin{theorem}
  Under conditions (C1)-(C4) given in Appendix, with the estimator $\widehat{\bm{\beta}}$ obtained from the estimation equation
  \begin{equation*}
    \sum_{i=1}^{n}\left[g(Y_i, \widehat{\bm{\beta}}^{T}\phi(x_i)) - \widehat{E}\{g(Y_i, \widehat{\bm{\beta}}^{T}\phi(x_i))|\widehat{\bm{\beta}}^{T}\phi(x_i)\}\right]\left[a(x_i) - \widehat{E}\{a(x_i)|\widehat{\bm{\beta}}^{T}\phi(x_i)\}\right] + n\widehat{\bm{\beta}}\bm{\Gamma} = 0
  \end{equation*}
  we have
  \begin{equation*}
    \left\|\widehat{\bm{\beta}} - \bm{\beta}\right\| \stackrel{Pr}\longrightarrow 0
  \end{equation*}
  where $\bm{\beta}$ is the true value and $\bm{\Gamma} = diag\{\gamma_1,\ldots,\gamma_q\}$.
\end{theorem}

\textbf{Proof for Theorem:}

First, we denote
  \begin{align*}
    \widehat{Q}_n(\widehat{\bm{\beta}}) & = \sum_{i=1}^{n}\left[g(Y_i, \widehat{\bm{\beta}}^{T}\phi(x_i)) - \widehat{E}\{g(Y_i, \widehat{\bm{\beta}}^{T}\phi(x_i))|\widehat{\bm{\beta}}^{T}\phi(x_i)\}\right]\left[a(x_i) - \widehat{E}\{a(x_i)|\widehat{\bm{\beta}}^{T}\phi(x_i)\}\right] \\
    Q_n(\bm{\beta}) & = \sum_{i=1}^{n}\left[g(Y_i, \bm{\beta}^{T}\phi(x_i)) - E\{g(Y_i, \bm{\beta}^{T}\phi(x_i))|\bm{\beta}^{T}\phi(x_i)\}\right]\left[a(x_i) - E\{a(x_i)|\bm{\beta}^{T}\phi(x_i)\}\right]
  \end{align*}
Rewrite $\widehat{Q}_n(\widehat{\bm{\beta}})$, we have
\begin{align*}
    & \quad \sum_{i=1}^{n}\left[g(Y_i, \widehat{\bm{\beta}}^{T}\phi(x_i)) - \widehat{E}\{g(Y_i, \widehat{\bm{\beta}}^{T}\phi(x_i))|\widehat{\bm{\beta}}^{T}\phi(x_i)\}\right]\left[a(x_i) - \widehat{E}\{a(x_i)|\widehat{\bm{\beta}}^{T}\phi(x_i)\}\right] \\
  = & \quad \sum_{i=1}^{n}\left[g(Y_i, \widehat{\bm{\beta}}^{T}\phi(x_i)) - E\{g(Y_i, \widehat{\bm{\beta}}^{T}\phi(x_i))|\widehat{\bm{\beta}}^{T}\phi(x_i)\}\right]\left[a(x_i) - E\{a(x_i)|\widehat{\bm{\beta}}^{T}\phi(x_i)\}\right] \\
  + & \quad \sum_{i=1}^{n}\left[g(Y_i, \widehat{\bm{\beta}}^{T}\phi(x_i)) - E\{g(Y_i, \widehat{\bm{\beta}}^{T}\phi(x_i))|\widehat{\bm{\beta}}^{T}\phi(x_i)\}\right]\left[E\{a(x_i)|\widehat{\bm{\beta}}^{T}\phi(x_i)\} - \widehat{E}\{a(x_i)|\widehat{\bm{\beta}}^{T}\phi(x_i)\}\right] \\
  + & \quad \sum_{i=1}^{n}\left[E\{g(Y_i, \widehat{\bm{\beta}}^{T}\phi(x_i))|\widehat{\bm{\beta}}^{T}\phi(x_i)\} - \widehat{E}\{g(Y_i, \widehat{\bm{\beta}}^{T}\phi(x_i))|\widehat{\bm{\beta}}^{T}\phi(x_i)\}\right]\left[a(x_i) - E\{a(x_i)|\widehat{\bm{\beta}}^{T}\phi(x_i)\}\right] \\
  + & \quad \sum_{i=1}^{n}\left[E\{g(Y_i, \widehat{\bm{\beta}}^{T}\phi(x_i))|\widehat{\bm{\beta}}^{T}\phi(x_i)\} - \widehat{E}\{g(Y_i, \widehat{\bm{\beta}}^{T}\phi(x_i))|\widehat{\bm{\beta}}^{T}\phi(x_i)\}\right]\left[E\{a(x_i)|\widehat{\bm{\beta}}^{T}\phi(x_i)\} - \widehat{E}\{a(x_i)|\widehat{\bm{\beta}}^{T}\phi(x_i)\}\right]\\
  = & \quad \bm{I}_1 + \bm{I}_2 + \bm{I}_3 + \bm{I}_4
\end{align*}

Using Taylor expansion, $\bm{I}_1$ can be converted as
\begin{equation*}
  \begin{split}
     & \quad \sum_{i=1}^{n}\left[g(Y_i, \widehat{\bm{\beta}}^{T}\phi(x_i)) - E\{g(Y_i, \widehat{\bm{\beta}}^{T}\phi(x_i))|\widehat{\bm{\beta}}^{T}\phi(x_i)\}\right]\left[a(x_i) - E\{a(x_i)|\widehat{\bm{\beta}}^{T}\phi(x_i)\}\right] \\
  = & \quad \sum_{i=1}^{n}\left[g(Y_i, \bm{\beta}^{T}\phi(x_i)) - E\{g(Y_i, \bm{\beta}^{T}\phi(x_i))|\bm{\beta}^{T}\phi(x_i)\}\right]\left[a(x_i) - E\{a(x_i)|\bm{\beta}^{T}\phi(x_i)\}\right] \\
  + & \quad nE\left\{\frac{\delta \left[g(Y, \bm{\beta}^{T}\phi(x)) - E\{g(Y, \bm{\beta}^{T}\phi(x))|\bm{\beta}^{T}\phi(x)\}\right]\left[a(x) - E\{a(x)|\bm{\beta}^{T}\phi(x)\}\right]}{\delta \bm{\beta}^{T}}\right\}\left(\widehat{\bm{\beta}} - \bm{\beta}\right) + \sum_{i=1}^{n}G_{i}(\bm{\beta},\widehat{\bm{\beta}} - \bm{\beta})
  \end{split}
\end{equation*}
where $\delta \left[g(Y, \bm{\beta}^{T}\phi(x)) - E\{g(Y, \bm{\beta}^{T}\phi(x))|\bm{\beta}^{T}\phi(x)\}\right]\left[a(x) - E\{a(x)|\bm{\beta}^{T}\phi(x)\}\right]\Big/\delta \bm{\beta}^{T}$ is the $Fr\acute{e}chet$ derivative.
we can set
$$
Q(\bm{\beta}) = \left[g(Y, \bm{\beta}^{T}\phi(x)) - E\{g(Y, \bm{\beta}^{T}\phi(x))|\bm{\beta}^{T}\phi(x)\}\right]\left[a(x) - E\{a(x)|\bm{\beta}^{T}\phi(x)\}\right]
$$
then
\begin{equation*}
  \delta Q(\bm{\beta})\Big/\delta \bm{\beta}^{T} =
  \begin{pmatrix}
    \delta Q_{1}(\bm{\beta})/\delta \bm{\beta}_{1} & \cdots & \delta Q_{1}(\bm{\beta})/\delta \bm{\beta}_{q} \\
    \vdots & \ddots & \vdots \\
    \delta Q_{q}(\bm{\beta})/\delta \bm{\beta}_{1} & \cdots & \delta Q_{q}(\bm{\beta})/\delta \bm{\beta}_{q}
  \end{pmatrix}
\end{equation*}
where
\begin{equation*}
  \delta Q_{i}(\bm{\beta})\Big/\delta \bm{\beta}_{j} =
  \begin{pmatrix}
    \partial \{Q_{i}(\bm{\beta})\}_{(1)}/\partial \{\bm{\beta}_{j}\}_{(1)} & \cdots & \{Q_{i}(\bm{\beta})\}_{(1)}/\partial \{\bm{\beta}_{j}\}_{(k)} & \cdots \\
    \vdots & \ddots & \vdots & \vdots\\
    \partial \{Q_{i}(\bm{\beta})\}_{(\ell)}/\partial \{\bm{\beta}_{j}\}_{(1)} & \cdots & \{Q_{i}(\bm{\beta})\}_{(\ell)}/\partial \{\bm{\beta}_{j}\}_{(k)} & \cdots\\
    \vdots & \vdots & \vdots & \vdots
  \end{pmatrix}
\end{equation*}

Moreover, we can set $A(x_0 + h)$ as $\left[g(Y_i, \widehat{\bm{\beta}}^{T}\phi(x_i)) - E\{g(Y_i, \widehat{\bm{\beta}}^{T}\phi(x_i))|\widehat{\bm{\beta}}^{T}\phi(x_i)\}\right]\left[a(x_i) - E\{a(x_i)|\widehat{\bm{\beta}}^{T}\phi(x_i)\}\right]$, then 
\begin{equation*}
  \left\|G_{i}(\bm{\beta},\widehat{\bm{\beta}} - \bm{\beta})\right\| \le \int_{0}^{1}\left\|\left[A^{\prime}\left(\bm{\beta} + t(\widehat{\bm{\beta}} - \bm{\beta})\right) - A^{\prime}(\bm{\beta})\right](\widehat{\bm{\beta}} - \bm{\beta})\right\|dt \le M_{1}\int_{0}^{1}t\|\widehat{\bm{\beta}} - \bm{\beta}\|^2dt = \frac{1}{2}M_1\|\widehat{\bm{\beta}} - \bm{\beta}\|^2
\end{equation*}
and the order of $G_{i}(\bm{\beta},\widehat{\bm{\beta}} - \bm{\beta})$ is $O_p(\rho_{n}^2)$. So the order of the remainder is $o_p(n^{1/2})$.

To deal with $\bm{I}_2$, we use Lemma \ref{lemmaA6.1} and Taylor expansions, then under the condition (C4), we have
\begin{equation*}
  \begin{split}
      & \quad \sum_{i=1}^{n}\left[g(Y_i, \widehat{\bm{\beta}}^{T}\phi(x_i)) - E\{g(Y_i, \widehat{\bm{\beta}}^{T}\phi(x_i))|\widehat{\bm{\beta}}^{T}\phi(x_i)\}\right]\left[E\{a(x_i)|\widehat{\bm{\beta}}^{T}\phi(x_i)\} - \widehat{E}\{a(x_i)|\widehat{\bm{\beta}}^{T}\phi(x_i)\}\right]\\
    = & \quad \sum_{i=1}^{n}\left[g(Y_i, \widehat{\bm{\beta}}^{T}\phi(x_i)) - E\{g(Y_i, \widehat{\bm{\beta}}^{T}\phi(x_i))|\widehat{\bm{\beta}}^{T}\phi(x_i)\}\right]\left[E\{a(x_i)|\bm{\beta}^{T}\phi(x_i)\} - \widehat{E}\{a(x_i)|\bm{\beta}^{T}\phi(x_i)\}\right]\left(1 + o_p(1)\right) \\
    = & \quad \sum_{i=1}^{n}\left[g(Y_i, \bm{\beta}^{T}\phi(x_i)) - E\{g(Y_i, \bm{\beta}^{T}\phi(x_i))|\bm{\beta}^{T}\phi(x_i)\}\right]\left[E\{a(x_i)|\bm{\beta}^{T}\phi(x_i)\} - \widehat{E}\{a(x_i)|\bm{\beta}^{T}\phi(x_i)\}\right]\left(1 + o_p(1)\right)
    \end{split}
\end{equation*}

Further, by Lemma \ref{lemmaA6.2}, the order of $\bm{I}_{2}$ is $o_p(n^{1/2})$ when $nh^{4m} \longrightarrow 0$ and $nh^{2d} \longrightarrow \infty$. Next, we focus on $\bm{I}_{3}$. With Lemma \ref{lemmaA6.1}, we can obtain that
\begin{equation}\label{A6.3.3}
  \begin{split}
     & \quad \sum_{i=1}^{n}\left[E\{g(Y_i, \widehat{\bm{\beta}}^{T}\phi(x_i))|\widehat{\bm{\beta}}^{T}\phi(x_i)\} - \widehat{E}\{g(Y_i, \widehat{\bm{\beta}}^{T}\phi(x_i))|\widehat{\bm{\beta}}^{T}\phi(x_i)\}\right]\left[a(x_i) - E\{a(x_i)|\widehat{\bm{\beta}}^{T}\phi(x_i)\}\right] \\
   = & \quad \sum_{i=1}^{n}\left[E\{g(Y_i, \bm{\beta}^{T}\phi(x_i))|\bm{\beta}^{T}\phi(x_i)\} - \widehat{E}\{g(Y_i, \bm{\beta}^{T}\phi(x_i))|\bm{\beta}^{T}\phi(x_i)\}\right]\left[a(x_i) - E\{a(x_i)|\widehat{\bm{\beta}}^{T}\phi(x_i)\}\right]\left(1 + o_p(1)\right) \\
   = & \quad \sum_{i=1}^{n}\left[E\{g(Y_i, \bm{\beta}^{T}\phi(x_i))|\bm{\beta}^{T}\phi(x_i)\} - \widehat{E}\{g(Y_i, \bm{\beta}^{T}\phi(x_i))|\bm{\beta}^{T}\phi(x_i)\}\right]\left[a(x_i) - E\{a(x_i)|\bm{\beta}^{T}\phi(x_i)\}\right]\left(1 + o_p(1)\right) \\
   + & \quad \sum_{i=1}^{n}\left[E\{g(Y_i, \bm{\beta}^{T}\phi(x_i))|\bm{\beta}^{T}\phi(x_i)\} - \widehat{E}\{g(Y_i, \bm{\beta}^{T}\phi(x_i))|\bm{\beta}^{T}\phi(x_i)\}\right]\left[E\{a(x_i)|\widehat{\bm{\beta}}^{T}\phi(x_i) - E\{a(x_i)|\bm{\beta}^{T}\phi(x_i)\}\right]
  \end{split}
\end{equation}

Then by Lemma \ref{lemmaA6.2}, the first term in the right side of (\ref{A6.3.3}) is of order $o_p(n^{1/2})$, and the second term is also of order $o_p(n^{1/2})$ while using the uniform convergence of nonparametric regression (Mack and Silverman, 1982) and Taylor expansion. Hence the order of (\ref{A6.3.3}) is $o_p(n^{1/2})$.

By the uniform convergence of nonparametric regression (Mack and Silverman, 1982), the order of $\bm{I}_4$ is $o_p(n^{1/2})$ when $nh^{4m} \longrightarrow 0$ and $nh^{2d} \longrightarrow \infty$.

Combining the above results for $\bm{I}_1$, $\bm{I}_2$, $\bm{I}_3$ and $\bm{I}_4$, we have
\begin{equation*}
  \sup_{\|\widehat{\bm{\beta}} - \bm{\beta}\| \le \rho_{n}}\|n^{-1/2}\widehat{Q}_n(\widehat{\bm{\beta}}) - n^{-1/2}Q_n(\bm{\beta}) - n^{1/2}A\left(\widehat{\bm{\beta}} - \bm{\beta}\right)\| = o_p(1)
\end{equation*}
where
\begin{equation*}
  A = E\left\{\frac{\delta \left[g(Y, \bm{\beta}^{T}\phi(x)) - E\{g(Y, \bm{\beta}^{T}\phi(x))|\bm{\beta}^{T}\phi(x)\}\right]\left[a(x) - E\{a(x)|\bm{\beta}^{T}\phi(x)\}\right]}{\delta \bm{\beta}^{T}}\right\}
\end{equation*}
To be convenient, we set $\gamma_{i} = \gamma_{n}, i= 1,2, \ldots, q$. Then we can obtain
\begin{align*}
       & \quad \inf_{\|\widehat{\bm{\beta}} - \bm{\beta}\| = \rho_{n}}\left\langle \widehat{\bm{\beta}} - \bm{\beta}, \widehat{Q}_n(\widehat{\bm{\beta}}) + n\widehat{\bm{\beta}}\bm{\Gamma} \right\rangle \\
     = & \quad \inf_{\|\widehat{\bm{\beta}} - \bm{\beta}\| = \rho_{n}}n^{1/2}\left\langle \widehat{\bm{\beta}} - \bm{\beta}, n^{-1/2}\widehat{Q}_n(\widehat{\bm{\beta}}) + n^{1/2}\widehat{\bm{\beta}}\bm{\Gamma} \right\rangle\\
     = & \quad \inf_{\|\widehat{\bm{\beta}} - \bm{\beta}\| = \rho_{n}}n^{1/2}\left\langle \widehat{\bm{\beta}} - \bm{\beta}, n^{-1/2}Q_n(\bm{\beta}) + n^{1/2}\bm{\beta}\bm{\Gamma} + n^{1/2}A\left(\widehat{\bm{\beta}} - \bm{\beta}\right) + n^{1/2}\left(\widehat{\bm{\beta}} - \bm{\beta}\right)\bm{\Gamma}  \right\rangle + o_p(1)\\
     = &  \quad \inf_{\|\widehat{\bm{\beta}} - \bm{\beta}\| = \rho_{n}}n\left\langle \widehat{\bm{\beta}} - \bm{\beta}, A\left(\widehat{\bm{\beta}} - \bm{\beta}\right) + \left(\widehat{\bm{\beta}} - \bm{\beta}\right)\bm{\Gamma}  \right\rangle + o_p(1)\\
     \ge & \quad (\lambda_{min}(A)+\gamma_{n})\rho_{n}^{2}n + o_p(1) \rightarrow \infty
\end{align*}
where
\begin{equation*}
  \|\widehat{\bm{\beta}} - \bm{\beta}\|^2 = \left\langle \widehat{\bm{\beta}} - \bm{\beta}, \widehat{\bm{\beta}} - \bm{\beta} \right\rangle = \sum_{j=1}^{q}\|\widehat{\bm{\beta}}_{j} - \bm{\beta}_{j}\|^2
\end{equation*}
Hence, we have
\begin{equation*}
  \lim_{n \rightarrow \infty}Pr\left(\inf_{\|\widehat{\bm{\beta}} - \bm{\beta}\| = \rho_{n}}\left\langle \widehat{\bm{\beta}} - \bm{\beta}, \widehat{Q}_n(\widehat{\bm{\beta}}) + n\gamma_n\widehat{\bm{\beta}} \right\rangle \ge 0 \right) = 1
\end{equation*}
then by using Leray-Schauder Theorem, we can obtain 
\begin{equation*}
  \left\|\widehat{\bm{\beta}} - \bm{\beta}\right\| \stackrel{Pr}\longrightarrow 0
\end{equation*}

\begin{theorem}
Suppose that the class of functions $\{\psi_{\theta,\eta}: \theta \in \Theta, \eta \in \mathcal{H}\}$ is P -Donsker, that the map $\theta \rightarrow P\psi_{\theta}$ is $Fr\acute{e}chet$ differentiable at $\theta_0$ with
derivative $A: lin\Theta \rightarrow \ell^{\infty}(\mathcal{H})$. Furthermore, assume that the maps $\theta \rightarrow \psi_{\theta, \eta}$ are continuous in $\ell_{2}(P)$ at $\theta_0$, uniformly in $\eta \in \mathcal{H}$. If $\lim_{n \rightarrow \infty}\sqrt{n}\gamma_{n} = 0$ and penalty function $q(\cdot)$ is bounded, Then any zero $\hat{\theta}_{n}$ of $\theta \rightarrow \mathbb{P}_{n}\psi_{\theta} + \gamma_{n}q(\theta)$ that converges in probability to a zero $\theta_{0}$ of $\theta \rightarrow P\psi_{\theta}$ satisfies
\begin{equation*}
  A\sqrt{n}(\hat{\theta}_{n} - \theta_{0}) = \mathbb{G}_{n}\psi_{\theta_{0}} + o_p(1)
\end{equation*}
\end{theorem}
\textbf{Proof:} Define the semi-metric of the space $\mathcal{H} \times \Theta$ by
\begin{equation*}
  d\left( (\eta, \theta), ( \eta^{\prime},  \theta^{\prime})\right) = \sqrt{P(\psi_{\eta, \theta} - \psi_{ \eta^{\prime},  \theta^{\prime}})^2}
\end{equation*}
and define a map $\phi: \ell^{\infty}(\mathcal{H} \times \Theta) \times \Theta \rightarrow \ell^{\infty}(\mathcal{H})$ by function $\phi(z,\theta) = z(\cdot, \theta) - z(\cdot, \theta_{0})$.

With the continuity of the maps $\theta \rightarrow \psi_{\theta, \eta}$, we have $d\left( (\eta, \theta), (\eta, \theta_{0})\right) \rightarrow 0 (\text{ as } \theta \rightarrow \theta_{0})$ uniformly in $\eta \in \mathcal{H}$. Therefor if $z \in \ell^{\infty}(\mathcal{H} \times \Theta)$ is d-uniformly continuous, we have $\left|z(\eta, \theta) - z(\eta, \theta_{0})\right| \rightarrow 0(\text{ as } \theta \rightarrow \theta_{0})$ uniformly in $\eta \in \mathcal{H}$.

Then for an arbitrary sequence in $\ell^{\infty}(\mathcal{H} \times \Theta) \times \Theta$ $(z_n, \theta_n) \rightarrow (z, \theta_0)$,
\begin{equation*}
  \begin{split}
     \|\phi(z_n, \theta_n) - \phi(z, \theta_0)\|_{\mathcal{H}} & = \|z_n(\eta, \theta_n) - z_n(\eta, \theta_0)\|_{\mathcal{H}} \\
       & = \|z_n(\eta, \theta_n) - z(\eta, \theta_n) + z(\eta, \theta_n) - z(\eta, \theta_0) + z(\eta, \theta_0) - z_n(\eta, \theta_0)\|_{\mathcal{H}} \\
       & \le 2\|z_n - z\|_{\mathcal{H} \times \Theta} + \|z(\eta, \theta_n) - z(\eta, \theta_0)\|_{\mathcal{H}} \rightarrow 0
  \end{split}
\end{equation*}
Hence, the map $\phi$ is continuous at every point $(z, \theta_{0})$. Because almost all sample paths of a Brownian bridge are uniformly continuous relative to the $\ell_{2}(P)$-norm, almost all sample paths $(\theta, \eta) \rightarrow Z(\theta, \eta)$ of the $Z(\theta, \eta) = \mathbb{G}\psi_{\theta,\eta}$ are uniformly
continuous relative to the metric d. With the boundness of $q(\cdot)$ and $\lim_{n \rightarrow \infty}\sqrt{n}\gamma_{n} = 0$, we have that $Z_n = \mathbb{G}_{n}\psi(\theta, \eta) + \sqrt{n}\gamma_{n}q_n(\theta) \rightsquigarrow Z$ and then $(Z_n, \hat{\theta}_{n}) \rightsquigarrow (Z, \theta_{0})$. Further, by the continuous mapping theorem, we obtain that $\phi(Z_n, \hat{\theta}_{n}) \rightsquigarrow \phi(Z, \theta_{0}) = 0$, which means
\begin{equation}\label{thm4_eq1}
  \mathbb{G}_{n}(\psi_{\hat{\theta}_{n}} - \psi_{\theta_{0}}) + \sqrt{n}\gamma_{n}\left(q(\hat{\theta}_{n}) - q(\hat{\theta}_{0})\right) \rightarrow^{P} 0
\end{equation}
As $\mathbb{P}_{n}\psi_{\hat{\theta}_{n}} + \gamma_{n}q(\hat{\theta}_{n}) = 0$ and $P\psi_{\theta_{0}} = 0$, (\ref{thm4_eq1}) can be converted as
\begin{equation*}
  -\mathbb{G}_{n}\psi_{\theta_{0}} - \sqrt{n}\gamma_{n}q(\theta_{0}) + \sqrt{n}\left(\mathbb{P}_{n}\psi_{\hat{\theta}_{n}} + \gamma_{n}q(\hat{\theta}_{n}) - P\psi_{\hat{\theta}_{n}} + P\psi_{\theta_{0}}\right) = o_p(1)
\end{equation*}
and
\begin{equation}\label{thm4_eq2}
  -\sqrt{n}\left(P\left(\psi_{\hat{\theta}_{n}} - \psi_{\theta_{0}}\right)\right) =  \mathbb{G}_{n}\psi_{\theta_{0}} + \sqrt{n}\gamma_{n}q(\theta_{0}) + o_p(1)
\end{equation}
By the assumption of Fr$\acute{e}$chet differentiability, (\ref{thm4_eq2}) can be written as

\begin{equation*}
  -\sqrt{n}A^{\star}\left(\hat{\theta}_{n} - \theta_{0}\right) = \mathbb{G}_{n}\psi_{\theta_{0}} + \sqrt{n}\gamma_{n}q(\theta_{0}) + o_p(1)
\end{equation*}
where $A^{\star}$ is the $Fr\acute{e}chet$ derivation of $P\psi_{\theta}$ at $\theta^{\star}$, $\theta^{\star}$ is between $\theta_{0}$ and $\hat{\theta}_{n}$.

As $\|A^{\star}_{(\theta^{\star})} - A_{(\theta_{0})}\| \le \|A\|\|\hat{\theta}_{n} - \theta_{0}\|$
and
\begin{equation*}
  \lim_{n \rightarrow \infty} Pr(\|\hat{\theta}_{n} - \theta_{0}\| \le \rho_{n}) = 1
\end{equation*}
where $\lim_{n \rightarrow \infty}\rho_{n} = 0$, then we have
\begin{equation*}
  \|A^{\star}_{(\theta^{\star})} - A_{(\theta_{0})}\| \rightarrow^{P} 0
\end{equation*}
Then we have
\begin{equation*}
  A\sqrt{n}(\hat{\theta}_{n} - \theta_{0}) = \mathbb{G}_{n}\psi_{\theta_{0}} + o_p(1)
\end{equation*}
and
\begin{equation*}
  A\sqrt{n}(\hat{\theta}_{n} - \theta_{0}) \rightsquigarrow \mathbb{G}\psi_{\theta_{0}}
\end{equation*}

\clearpage
\textbf{A.8 Figures of Simulation Studies}
\begin{figure}[ht]
    \centering
    \subfigure[p = 10]{
    \begin{minipage}[t]{0.5\linewidth}
        \centering
        \includegraphics[width=1\linewidth]{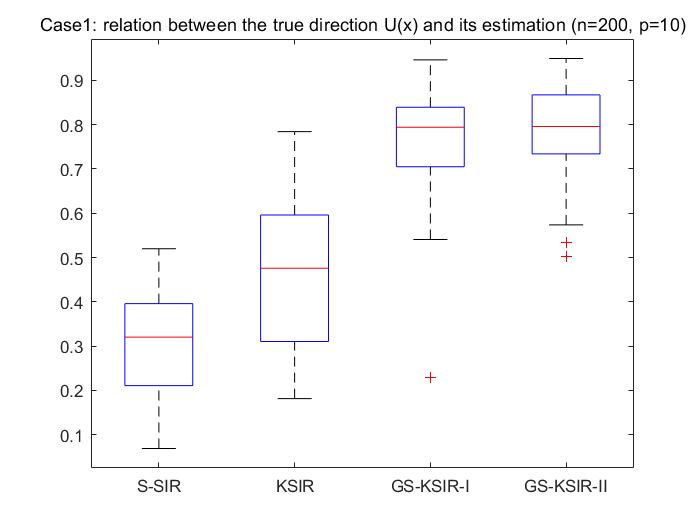}
    \end{minipage}%
    }%
    \subfigure[p = 20]{
    \begin{minipage}[t]{0.5\linewidth}
        \centering
        \includegraphics[width=1\linewidth]{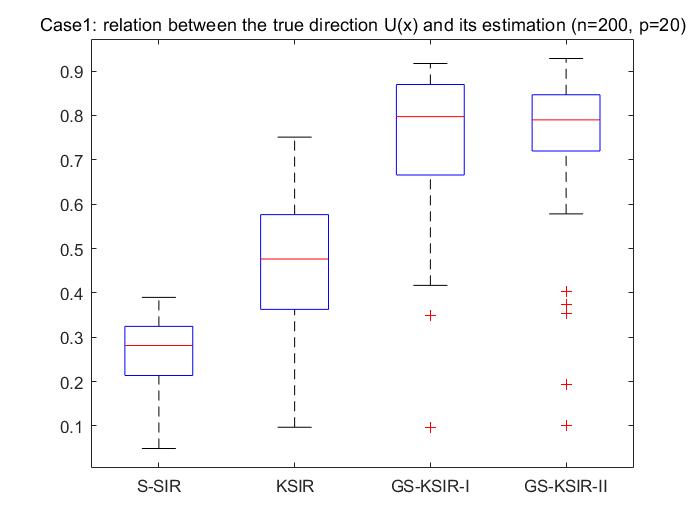}
    \end{minipage}%
    }%
    \centering
    \caption{Case 1 with p = 10 and p = 20}
\end{figure}

\begin{figure}[ht]
    \centering
    \subfigure[p = 10]{
    \begin{minipage}[t]{0.5\linewidth}
        \centering
        \includegraphics[width=1\linewidth]{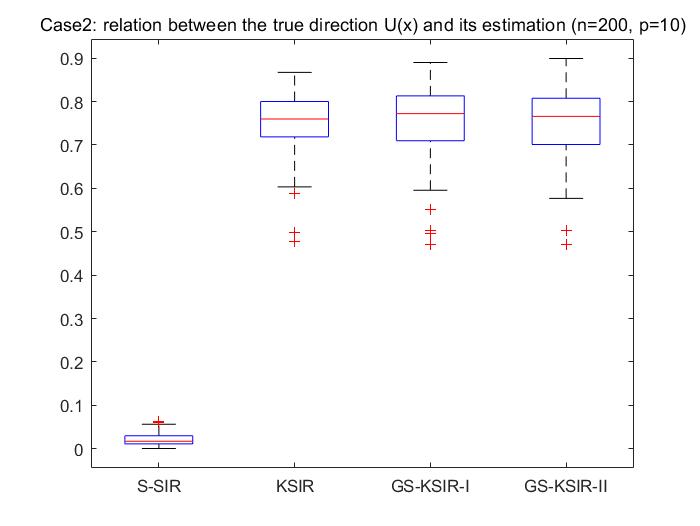}
    \end{minipage}%
    }%
    \subfigure[p = 20]{
    \begin{minipage}[t]{0.5\linewidth}
        \centering
        \includegraphics[width=1\linewidth]{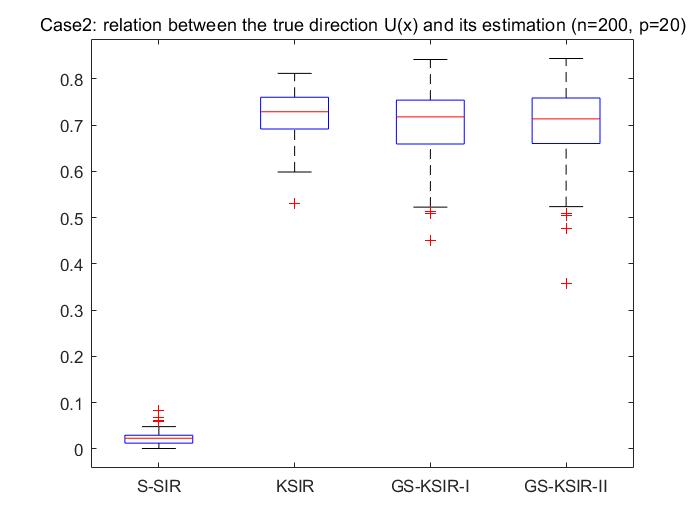}
    \end{minipage}%
    }%
    \centering
    \caption{Case 2 with p = 10 and p = 20}
\end{figure}

\begin{figure}[ht]
    \centering
    \subfigure[p = 10]{
    \begin{minipage}[t]{0.5\linewidth}
        \centering
        \includegraphics[width=1\linewidth]{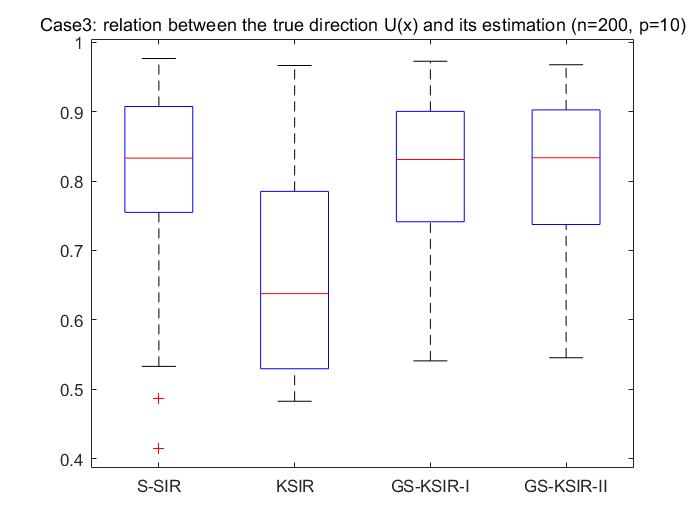}
    \end{minipage}%
    }%
    \subfigure[p = 20]{
    \begin{minipage}[t]{0.5\linewidth}
        \centering
        \includegraphics[width=1\linewidth]{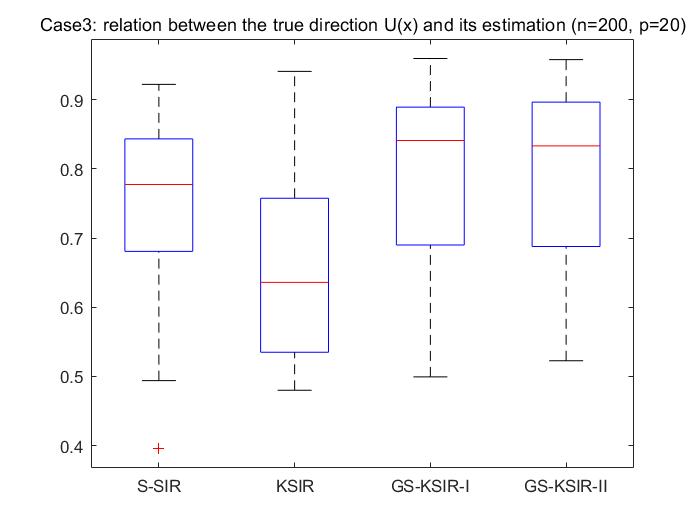}
    \end{minipage}%
    }%
    \centering
    \caption{Case 3 with p = 10 and p = 20}
\end{figure}

\end{document}